\documentclass[fleqn,10pt]{article}
\usepackage{epsfig}
\newcommand{\eq}[1]{eq.~(\ref{#1})} 
\newcommand{\be}{\begin{equation}}
\newcommand{\ee}{\end{equation}}
\newcommand{\alphabold}{\mbox{\small\boldmath $\alpha$}}
\newcommand{\xbold}{\mbox{\boldmath $x$}}

\newcommand{\delchisq}{\Delta \chi^2_i(x_i;\alphabold)}
\newcommand{\delchi}{\Delta \chi^2_i}
\newcommand{\delchisqmax}{{\Delta \chi^2_i(x_i;\alphabold)}_{\rm max}}
\newcommand{\delchimax}{{\delchi}_{\rm max}}
\newcommand{\x}{\nu/m)}

\newcommand{\rnd}{{\rm RND}}

\begin{document}
\renewcommand\thepage{\ }
\begin{titlepage} 
%
\newcommand\reportnumber{1010} 
\newcommand\mydate{May 2, 2005} 
\newlength{\nulogo} 
\settowidth{\nulogo}{\small\sf{NUHEP Report XXXX}}
\title{
\vspace{-.8in} 
\hfill\fbox{{\parbox{\nulogo}{\small\sf{
NUHEP Report  \reportnumber\\
          \mydate}}}}
\vspace{0.5in} \\
{
Sifting data in the real world
}}

\author{
M.~M.~Block\\
{\small\em Department of Physics and Astronomy,} \vspace{-5pt} \\ 
{\small\em Northwestern University, Evanston, IL 60208}\\
\vspace{-5pt}
\  \\
\vspace{-5pt}\\
%
\vspace{-5pt}\\
%
}    
\vspace{.5in}
\vfill
\date {}
\maketitle
\begin{abstract}
In the real world, experimental data are rarely, if ever, distributed as a normal (Gaussian) distribution. As an example, a large set of data---such as the cross sections for particle scattering as a function of energy contained in the archives of the Particle Data Group\cite{pdg}---is a compendium of all published data, and hence, unscreened. Inspection of similar data sets quickly shows that, for many reasons, these data sets have many outliers---points well beyond what is expected from a normal distribution---thus ruling out the use of conventional $\chi^2$ techniques. This note suggests an adaptive algorithm that allows a phenomenologist to apply to the data sample a sieve whose mesh is coarse enough to let the background fall through, but fine enough to retain the preponderance of the signal, thus sifting the data.  A prescription is given for  finding a robust estimate of the best-fit model parameters in the presence of a noisy background, together with a robust estimate of the model parameter errors,  as well as a determination of  the goodness-of-fit of the data to the theoretical hypothesis. Extensive computer simulations are carried out to test the algorithm for both its accuracy and stability under varying background conditions.
\end{abstract}
\end{titlepage} 
%
\pagenumbering{arabic}
\renewcommand{\thepage}{-- \arabic{page}\ --}  
\section{Introduction}
In an idealized world where all of the data follow a normal (Gaussian) distribution, the use of the $\chi^2$ likelihood technique, through minimization of $\chi^2$, described in detail in \ref{appendix:chisq}, offers a powerful statistical analysis tool when fitting  models to a data sample.
It allows the phenomenologist to conclude either:
\begin{itemize}
\item The model is accepted, based on the value of its $\chi^2_{\rm min}$. It certainly fits well when $\chi^2_{\rm min}$, when compared to $\nu$, the numbers of degrees of freedom, has a reasonably high probability ($\chi^2_{\rm min}\sim \nu$). 
On the other hand, it might be accepted with a much poorer $\chi^2_{\rm min}$, depending on the phenomenologist's judgment. In any event, the goodness-of-fit of the data to the model is known and an informed judgment can be made.
\item  Its parameter errors are such that  a change of $\Delta \chi^2=1$ from  $\chi^2_{\rm min}$ corresponds to changing a parameter by its standard error $\sigma$. These errors and their correlations are summarized in the standard covariance matrix $C$ discussed in Appendix \ref{appendix:chisq}.
\end{itemize}
or
\begin {itemize}
\item  The model is rejected, because  the probability that the data set fits the model  is too low, {\em i.e.,} $\chi^2_{\rm min}>>\nu$.
\end{itemize}
This decision-making capability (of accepting or rejecting the model) is of primary importance, as is the ability to estimate the parameter errors and their correlations.

Unfortunately, in the real world, experimental data sets are at best only approximately Gaussian and often are riddled with outliers---points far off from a best fit curve to the data, being many standard deviations away. This can be due to many sources, copying errors, bad measurements, wrong calibrations, misassignment of experimental errors, {\em etc.} It is this world that our note wishes to address---a world with many data points, and perhaps, many different experiments from many different experimenters, with possibly a significant number of  outliers. 

\begin{figure}[h,t,b,p] 
\begin{center}
\mbox{\epsfig{file=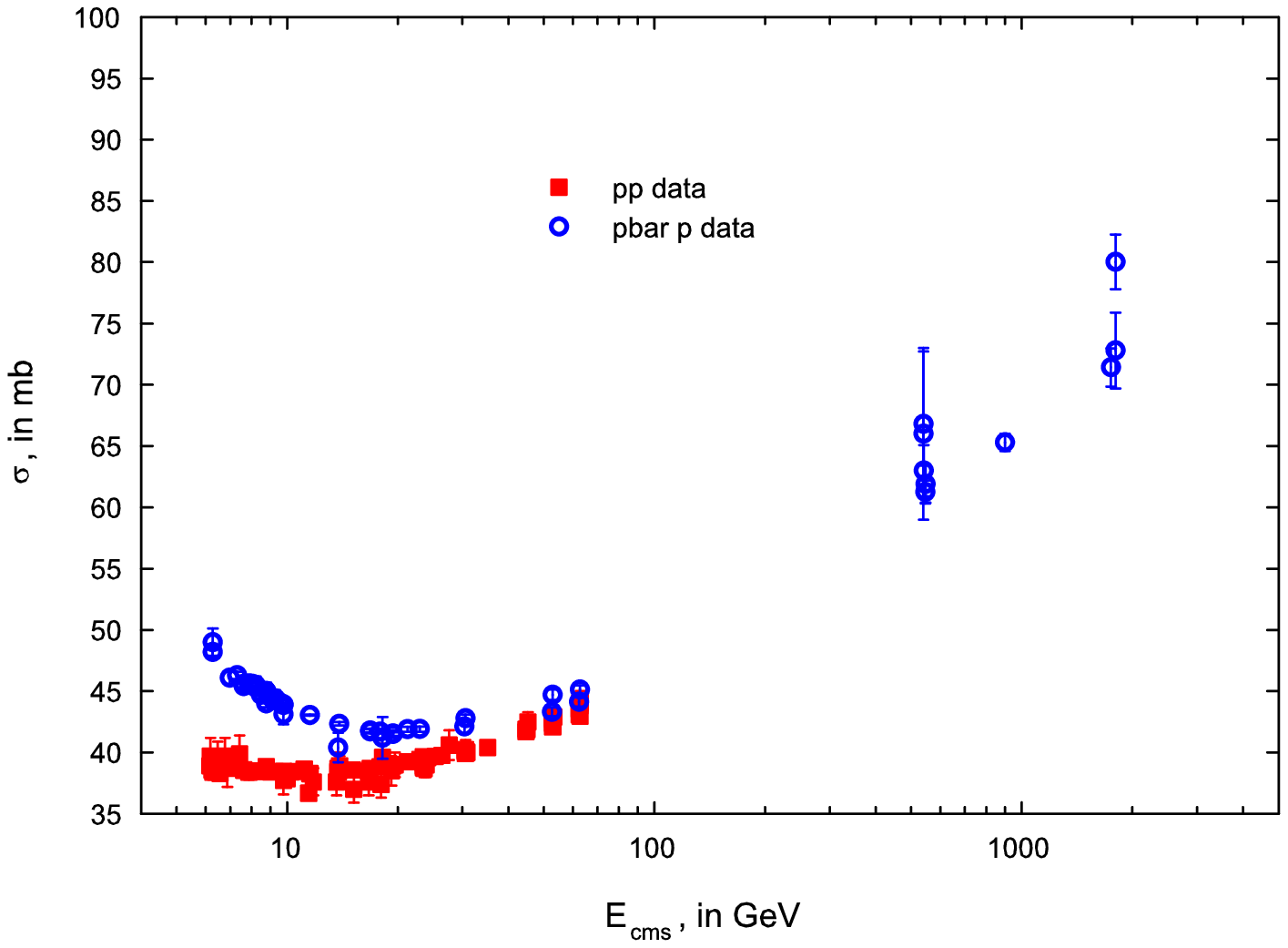,width=5in%
,bbllx=0pt,bblly=0pt,bburx=420pt,bbury=305pt,clip=%
}}
\end{center}
\caption[]{ \footnotesize
The data points shown are {\em all} of the experimental data listed in the Particle Data Group\cite{pdg} site for $\bar pp$ and $pp$ total cross sections in the energy interval ${\rm E}_{\rm cms}> 6$ GeV.  The open circles are $\sigma_{\bar pp}$ and the squares are $\sigma_{ pp}$.\ \ \ \ \ \ \ \ \ \ \
  }
\label{sigpdg}
\end{figure}

\begin{figure}[h,t,b,p] 
\begin{center}
\mbox{\epsfig{file=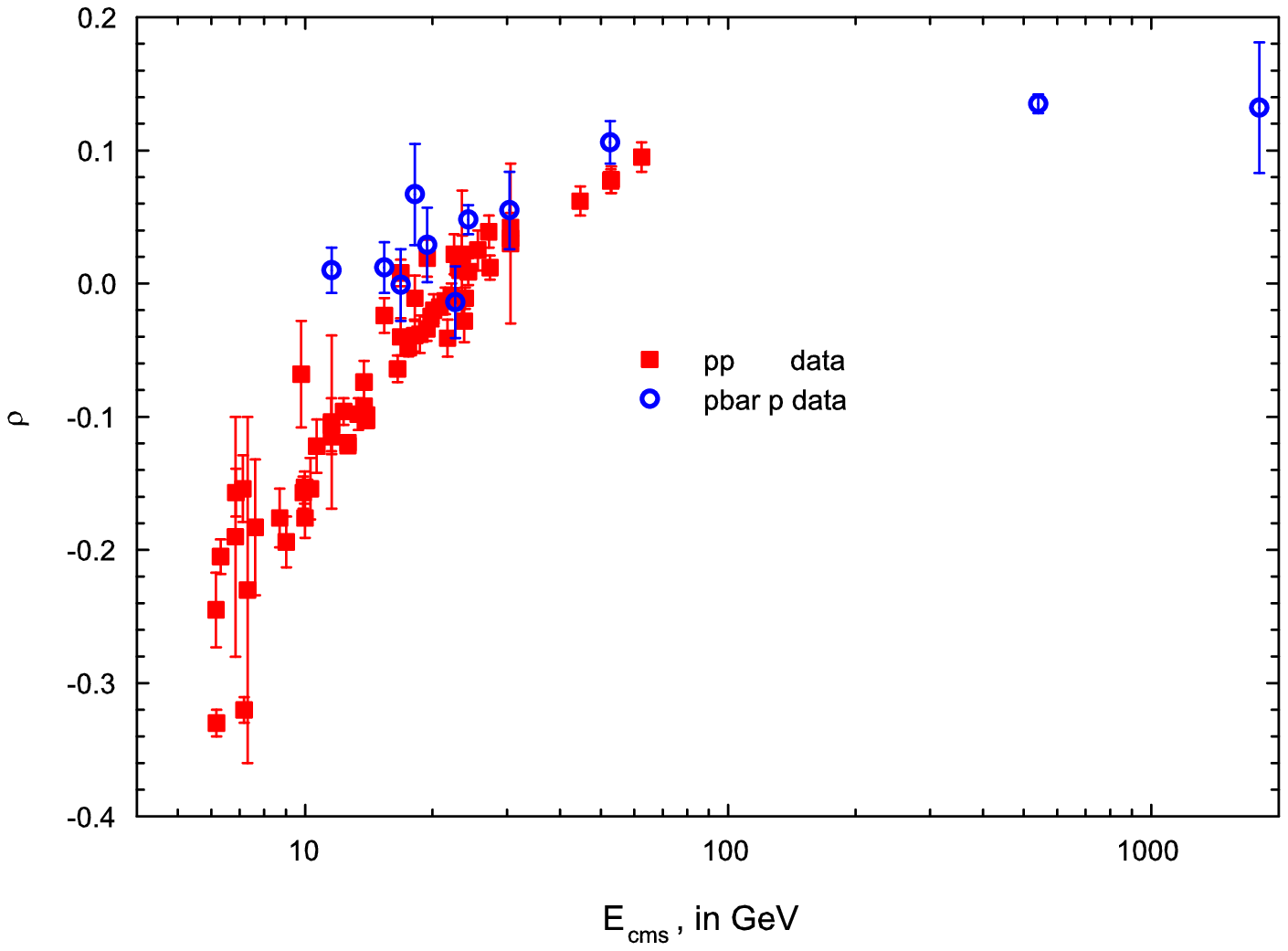,width=5in%
,bbllx=0pt,bblly=0pt,bburx=420pt,bbury=305pt,clip=%
}}
\end{center}
\caption[]{ \footnotesize
The data points shown are {\em all} of the experimental data listed in the Particle Data Group\cite{pdg} site for  $\bar pp$ and $pp$ $\rho$-values (ratio of the real to the imaginary portion of the forward scattering amplitude) in the energy interval ${\rm E}_{\rm cms} >6\  {\rm GeV}$.  The open circles are $\rho_{\bar pp}$ and the squares are $\rho_{ pp}$.
  }
\label{rhopdg}
\end{figure}

\begin{figure}[h,t,b,p] 
\begin{center}
\mbox{\epsfig{file=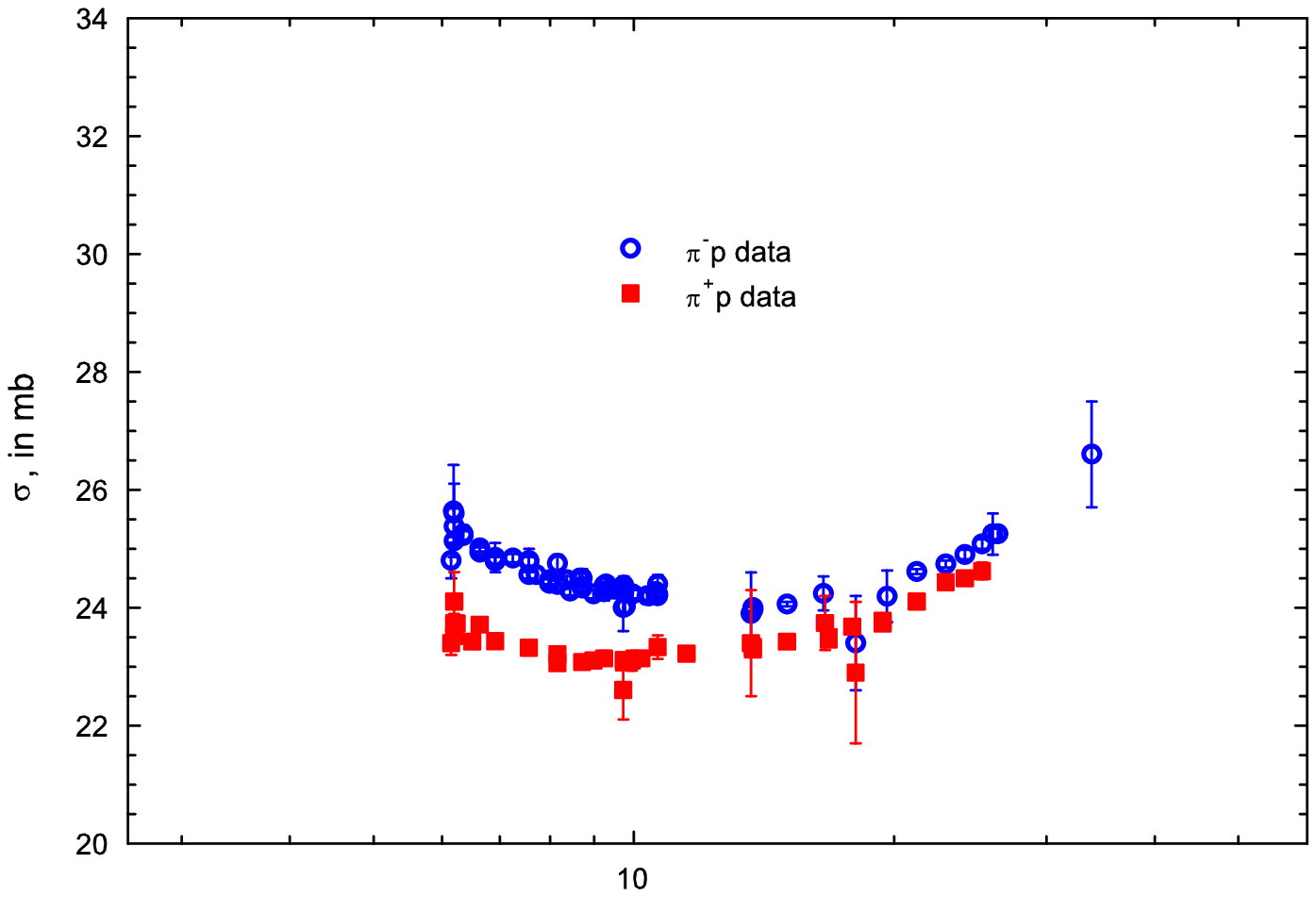,width=5in%
,bbllx=0pt,bblly=0pt,bburx=408pt,bbury=288pt,clip=%
}}
\end{center}
\caption[]{ \footnotesize
The data points shown are {\em all} of the experimental data listed in the Particle Data Group\cite{pdg} site for $\pi^-p$ and $\pi^+p$ total cross sections in the energy interval ${\rm E}_{\rm cms}> 6$ GeV.  The open circles are $\sigma_{\pi^- p}$ and the squares are $\sigma_{\pi^+p}$.
  }
\label{pipdg}
\end{figure}

\begin{figure}[h,t,b,p] 
\begin{center}
\mbox{\epsfig{file=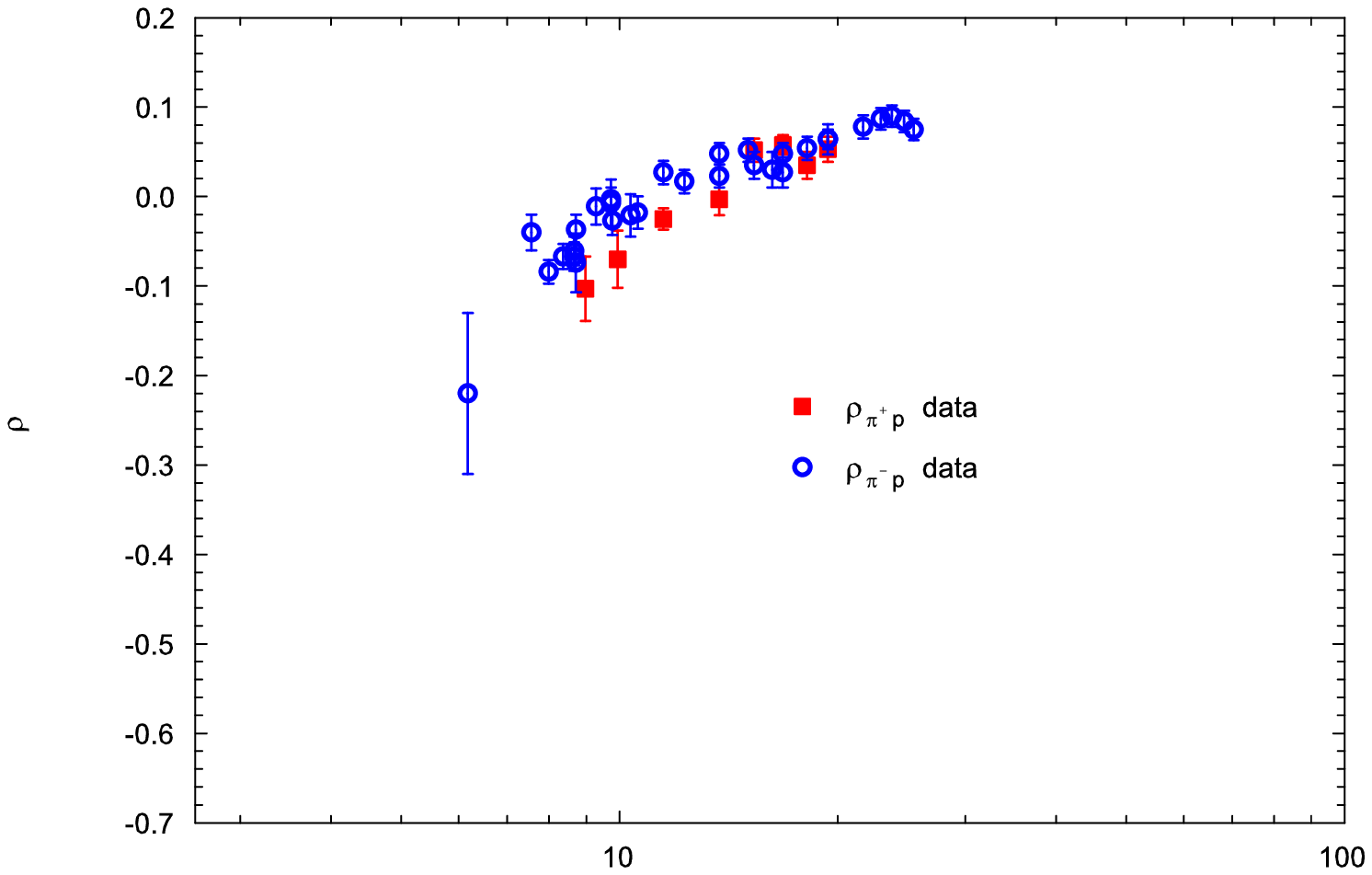,width=5.5in%
,bbllx=0pt,bblly=0pt,bburx=447pt,bbury=288pt,clip=%
}}
\end{center}
\caption[]{ \footnotesize
The data points shown are {\em all} of the experimental data listed in the Particle Data Group\cite{pdg} site for $\pi^-p$ and $\pi^+p$ $\rho$-values (ratio of the real to the imaginary portion of the forward scattering amplitude) in the energy interval ${\rm E}_{\rm cms} >6\  {\rm GeV}$.  The open circles are $\rho_{\pi^+ p}$ and the squares are $\rho_{\pi^-p}$.
  }
\label{pirhopdg}
\end{figure}

\begin{figure}[h,t,b,p] 
\begin{center}
\mbox{\epsfig{file=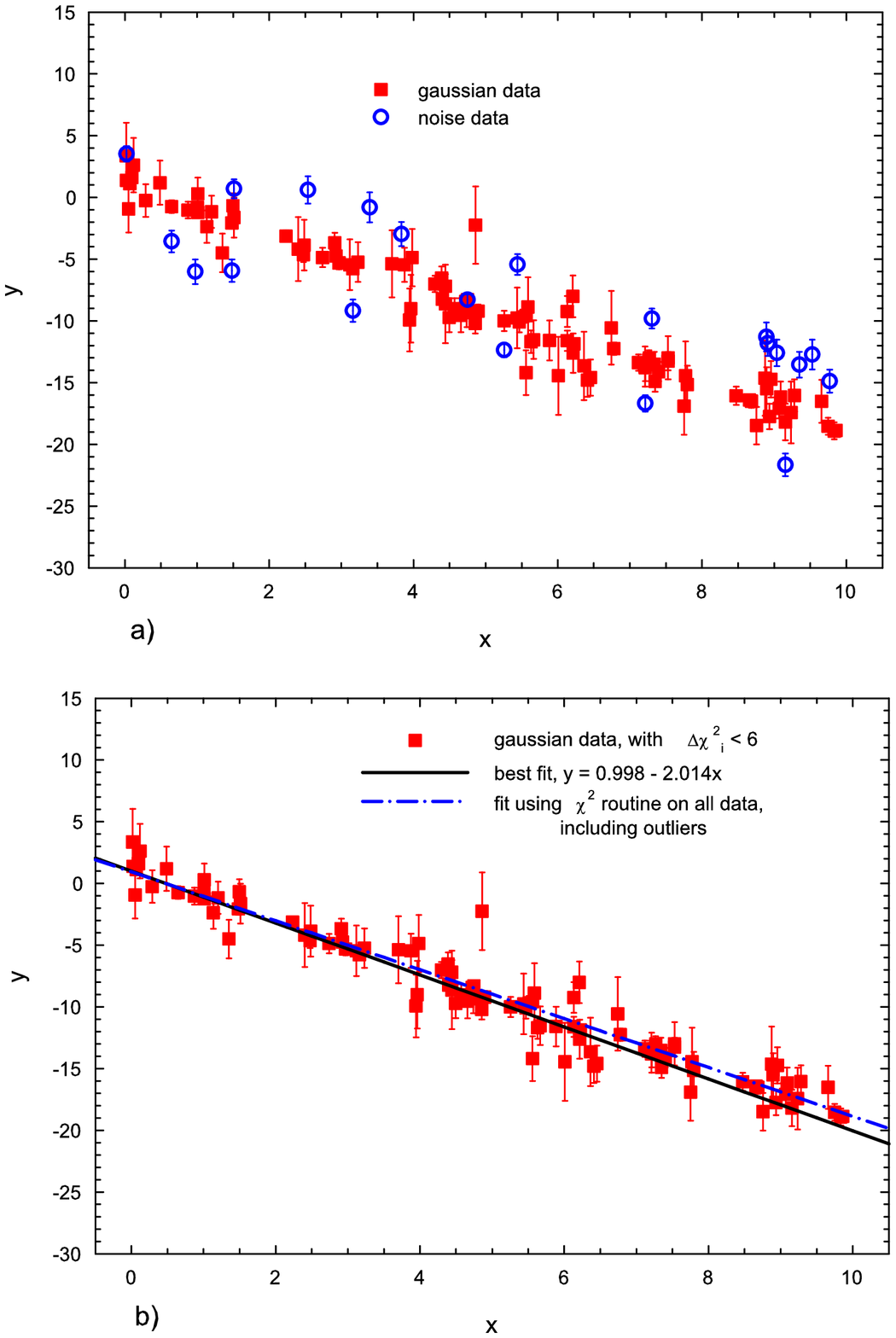,width=5in%
,bbllx=0pt,bblly=0pt,bburx=425pt,bbury=618pt,clip=%
}}
\end{center}
\caption[]{ \footnotesize
a) The 100 squares are a computer-generated Gaussianly distributed data set  about the straight line $y=1-2x$. The 20 open circles are randomly distributed noise data. See Section \ref{section:line} for details.

b) The 100 data points shown  are the result of screening all 120 data points  for those points having $\Delta \chi^2_i<6.$ 
There were no noise points (open circles)  retained in the Sieve and the 100 squares are the Gaussian data retained in the Sieve.
The best fit curve to all points with $\delchi <6$, $y=a+bx$, is the solid curve, where $a=0.998\pm0.12$, $b=-2.014\pm0.020$, and $\chi^2_{\rm min}/\nu=0.91$, yielding a renormalized value ${\cal R}\times\chi^2_{\rm min}/\nu=1.01$ compared to the expected $<\chi^2>/\nu=1.0\pm0.14$. The dashed-dot  curve  is a $\chi^2$ fit to the totality of data---100 signal plus 20 noise points---which has  $\chi^2_{\rm min}/\nu=4.8$. 
 }
\label{noisy}
\end{figure}

\begin{figure}[h,t,b,p] 
\begin{center}
\mbox{\epsfig{file=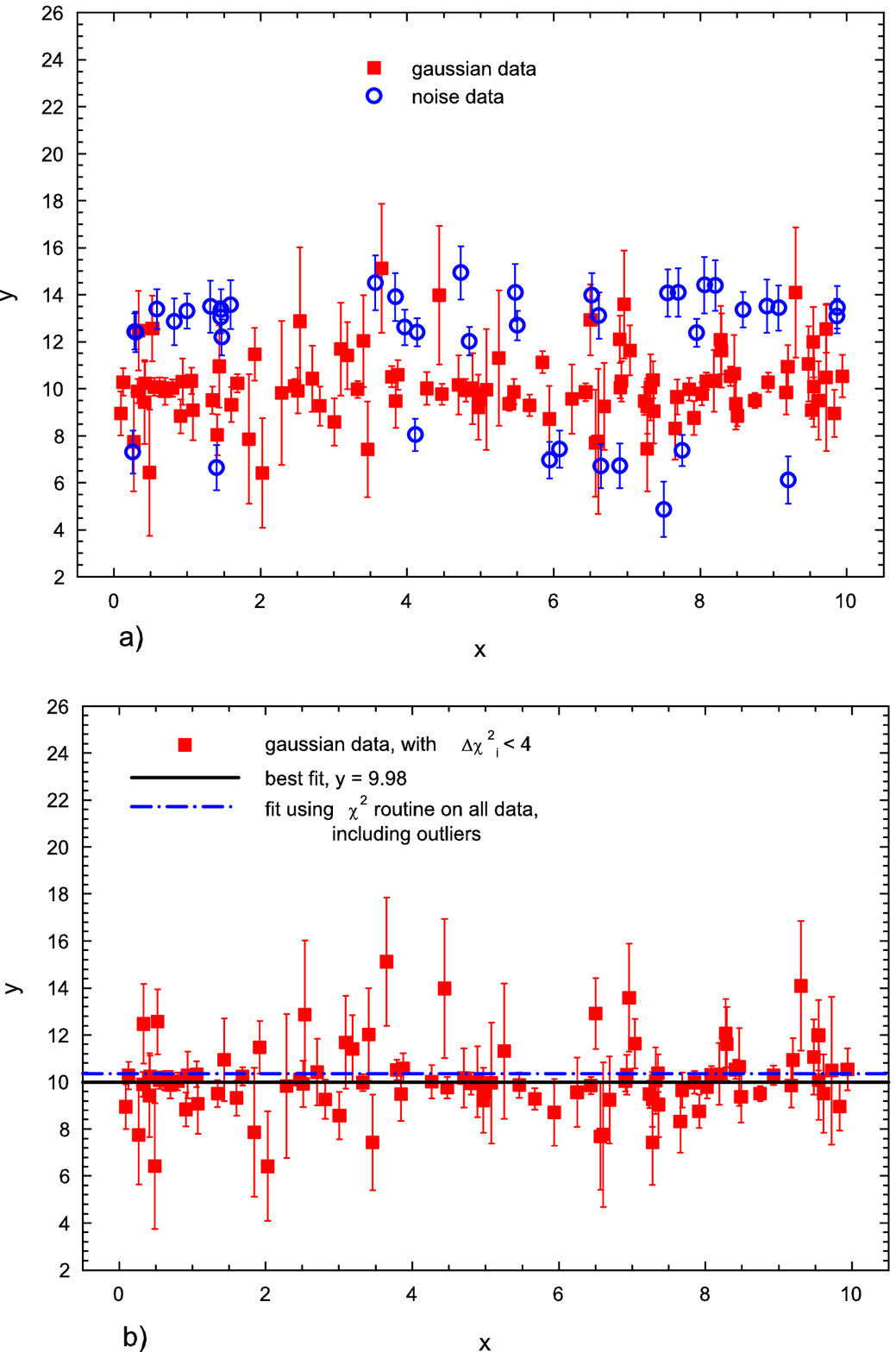,width=5in%
}}
\end{center}
\caption[]{ \footnotesize
a) The 100 squares are a computer-generated Gaussianly distributed data set  about the constant $y=10$. The 40 open circles are randomly distributed noise data. See Section \ref{section:constant} for details.

b) The 98 data points shown  are the result of screening all 140 data points  for those points having $\Delta \chi^2_i<4.$ 
There were no noise points (open circles)  retained in the Sieve and the 98 squares are the Gaussian data retained in the Sieve.
The best fit curve to all points with $\delchi <4$, $y=c$, is the solid curve, where $c=9.98\pm0.074$,  and $\chi^2_{\rm min}/\nu=0.84$, yielding a renormalized value ${\cal R}\times\chi^2_{\rm min}/\nu=1.09$  compared to the expected $<\chi^2>/\nu=1.0\pm0.14$. The dashed-dot  curve  is a $\chi^2$ fit to the totality of data---100 signal plus 40 noise points---which has  $\chi^2_{\rm min}/\nu=4.39$. 
 }
\label{constant140_4}
\end{figure}

\begin{figure}[h,t,b,p] 
\begin{center}
\mbox{\epsfig{file=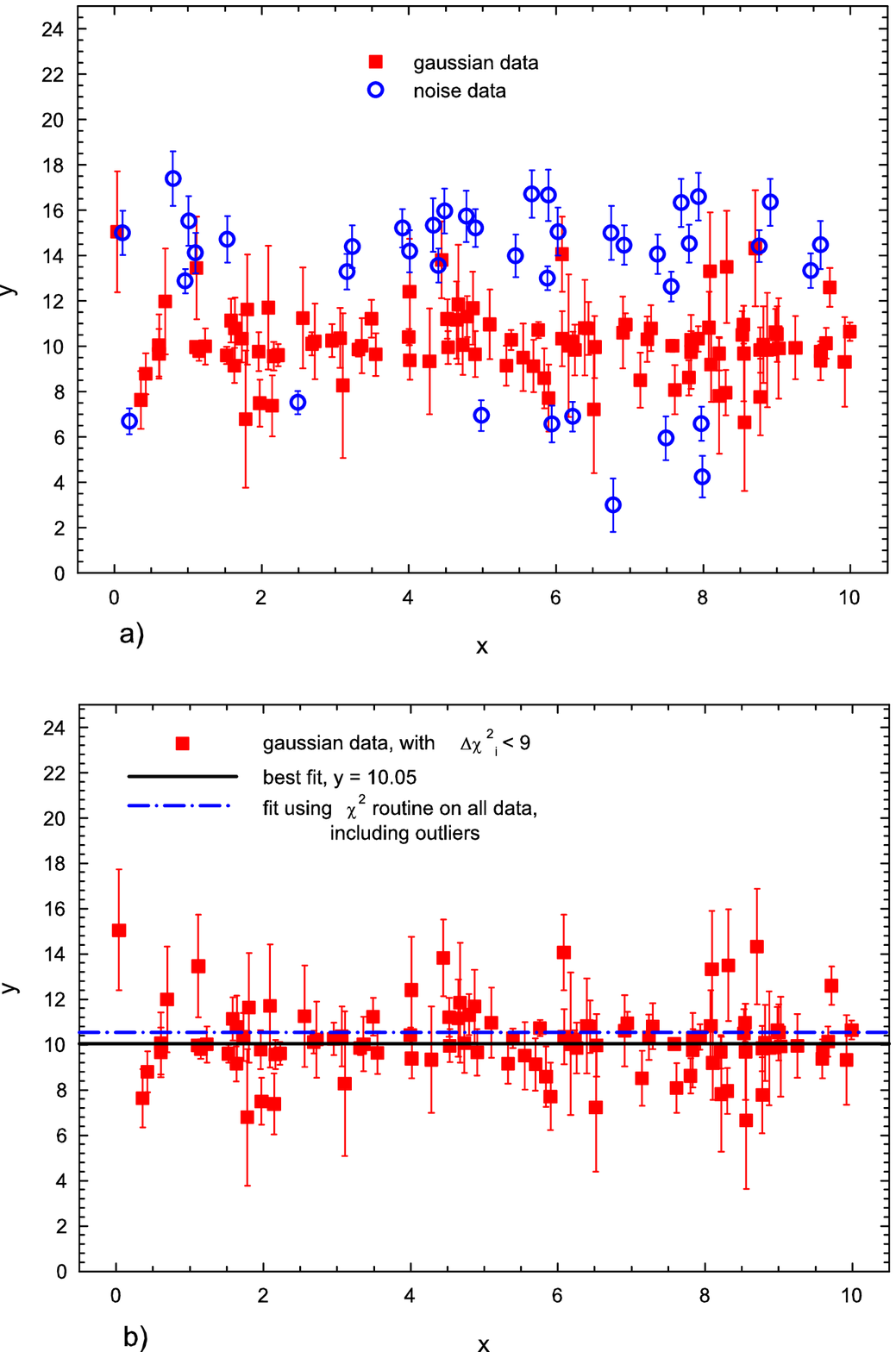,width=5in%
}}
\end{center}
\caption[]{ \footnotesize
a) The 100 squares are a computer-generated Gaussianly distributed data set  about the constant $y=10$. The 40 open circles are randomly distributed noise data. See Section \ref{section:constant} for details.

b) The 99 data points shown  are the result of screening all 140 data points  for those points having $\Delta \chi^2_i<9.$ 
There were no noise points (open circles)  retained in the Sieve and the 98 squares are the Gaussian data retained in the Sieve.
The best fit curve to all points with $\delchi <9$, $y=c$, is the solid curve, where $c=10.05\pm0.074$,  and $\chi^2_{\rm min}/\nu=1.08$, yielding a renormalized value ${\cal R}\times\chi^2_{\rm min}/\nu=1.11$ compared to the expected $<\chi^2>/\nu=1.0\pm0.14$. The dashed-dot  curve  is a $\chi^2$ fit to the totality of data---100 signal plus 40 noise points---which has  $\chi^2_{\rm min}/\nu=8.10$. 
 }
\label{constant140}
\end{figure}

\begin{figure}[h,t,b,p] 
\begin{center}
\mbox{\epsfig{file=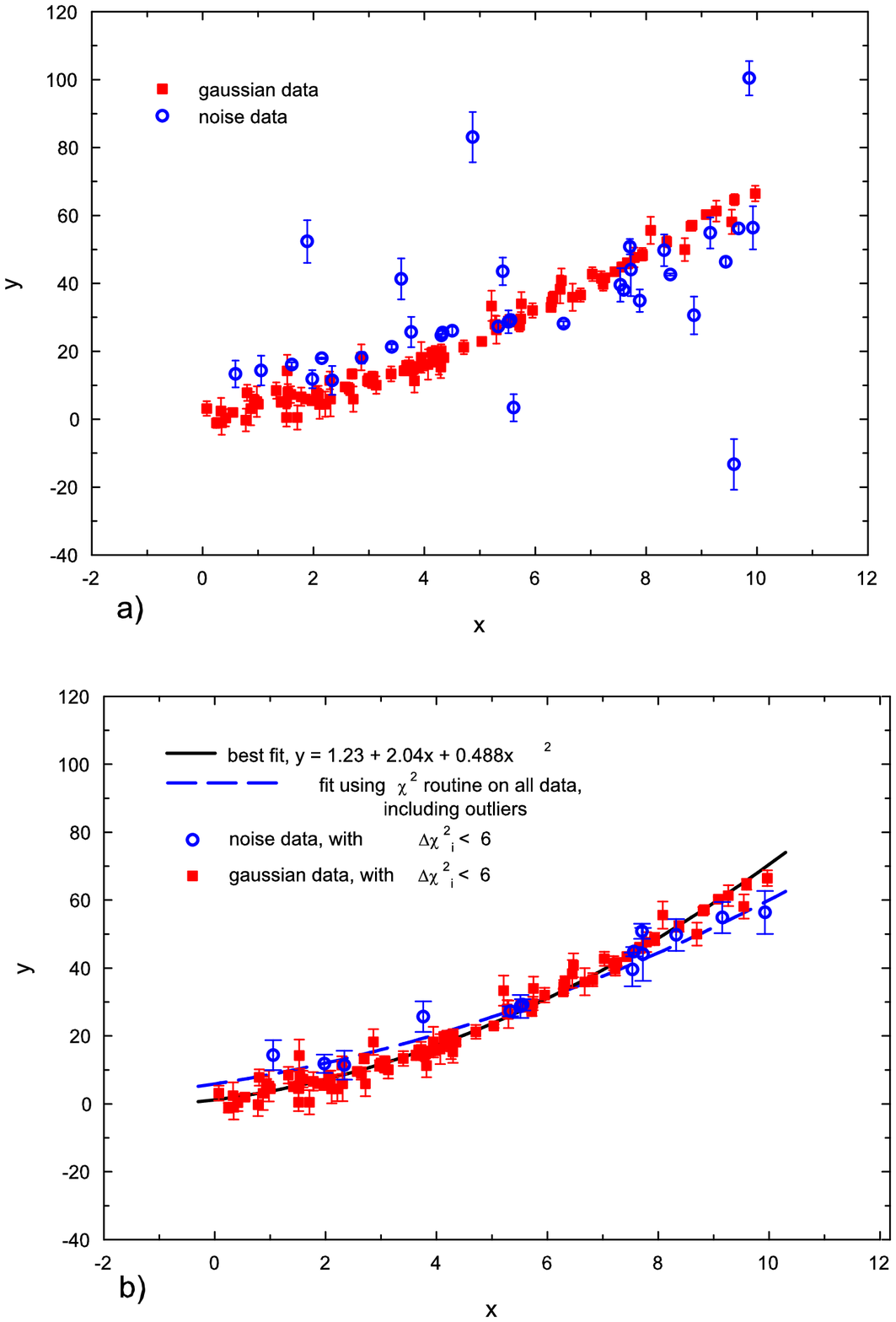,width=5.4in%
,bbllx=0pt,bblly=0pt,bburx=427pt,bbury=622pt,clip=%
}}
\end{center}
\caption[]{ \footnotesize
a) The 100 squares are a computer-generated Gaussianly distributed data set  about the parabola $y=1+2x+0.5x^2$. The 35 open circles are randomly distributed noise data around the parabola $y=12+2x+0.2x^2$. See Section \ref{section:parabola} for details.

b) The 113 data points shown  are the result of screening all of the data for those points having $\Delta \chi^2_i<6.$ 
The open circles are the 13 noise points retained in the Sieve and the 100 squares are the Gaussian data retained in the Sieve.
The best fit curve to all points with $\delchi <6$, $y=1.23+2.04x+0.48x^2$, is the solid curve. The dashed curve is a $\chi^2$ fit to the totality of data in Fig. \ref{noisyparabola}, consisting of signal plus noise.
  }
\label{noisyparabola}
\end{figure}

In Section \ref{section:sieve} we will propose our ``Sieve'' algorithm, an adaptive technique for discarding outliers while retaining the vast majority of the good data.
This then allows us to estimate the goodness-of fit and make a  robust determination of both the parameters and their errors---for a discussion of the term ``robust'', see Appendix {\ref{appendix}. In essence, we then retain all of the statistical benefits of the conventional $\chi^2$ technique.   

In Sections \ref{section:pp} and \ref{section:pip} we will apply the algorithm to high energy $\bar pp$ and $pp$ scattering, as well as to $\pi^- p$ and $\pi^+ p$ scattering. 
Eight examples of real world experimental data, for both $\bar pp$ and $pp$ scattering and $\pi^+p$ and $\pi^-p$ scattering, are taken from the Particle Data Group archives\cite{pdg} and  are  illustrated  in Figures  \ref{sigpdg},   \ref{rhopdg},  \ref{pipdg} and \ref{pirhopdg}, respectively. The data in Fig. \ref{sigpdg} are all of the known published data for the total cross sections $\sigma_{\bar pp}$ and 
$\sigma_{pp}$ for cms (center of mass) energies greater than 6 GeV. The measured $\rho_{\bar pp}$ and 
$\rho_{pp}$, where $\rho$ is the ratio of the real to the imaginary portion of the forward scattering amplitude, are shown in  Fig. \ref{rhopdg},  again for cms energies greater that 6 GeV. The data in Fig. \ref{pipdg} are all of the known published data for the total cross sections $\sigma_{\pi^-p}$ and 
$\sigma_{\pi^+p}$ for cms  energies greater that 6 GeV. The measured $\rho_{\pi^-p}$ and 
$\rho_{\pi^+p}$ are shown in  Fig. \ref{pirhopdg},  again for cms energies greater that 6 GeV.  Detailed examination of Figures \ref{sigpdg}, \ref{rhopdg}, \ref{pipdg} and  \ref{pirhopdg} show many points far off of the common trend, often at the same energy. Attempts to use the $\chi^2$ technique to fit these data with a model will always come up short. These fits will always return a huge value of $\chi^2_{\rm min}/\nu$, together with model parameters that are likely to be  unreliable.

In Section \ref{section:computer},  we make  three  types of computer simulations, generating data normally distributed about a straight line, a constant, and about a parabola, along with outliers---artificial worlds where we know all of the answers, {\em i.e.,} which points are signal and which are noise. Examples for the straight line, a constant (two cases) and the parabola are shown in Fig. \ref{noisy}, \ref{constant140_4}, \ref{constant140} and  \ref{noisyparabola}. Details are given in Sections \ref{section:line}, \ref{section:constant} and \ref{section:parabola}. The noise points in Fig. \ref{noisy}a, \ref{constant140_4}a,  \ref{constant140}a,  and Fig. \ref{noisyparabola}a are the diamonds, whereas  the signal points are the circles.  

The dashed curve in Fig. \ref{noisy}b is the result of a $\chi^2$ fit to all of the noisy data (100 signal plus 20 noise points) in Figure \ref{noisy}a and is not a very good  fit to the data. The solid line is the fit with the `` Sieve'' algorithm proposed in the next Section. It reproduces nicely the theoretical straight line $y=1-2x$ that was used to computer-generate data that were normally distributed about it, using random numbers. In this case, the 20 noise points penetrated the signal down to a level $\delchi>6$.

In Fig. \ref{constant140_4}b we show the results for fitting the constant $y=10$. The noise points (diamonds) in Fig. \ref{constant140_4}a penetrate the signal down to  $\delchi>4$. 

In Fig. \ref{constant140}b we show the results for fitting the constant $y=10$, where the noise points (diamonds) in Fig. \ref{constant140}a penetrate the signal down to  $\delchi>9$.  

In Fig. \ref{noisyparabola}a the data were generated about the parabola $y=1+2x+0.5x^2$, with background noise. Figure \ref{noisyparabola}b shows the result of sifting the data according to our Sieve algorithm, described below.   The noise points that are retained after invoking our algorithm are the diamonds in  Fig. \ref{noisyparabola}b and the circles are the signal points that are retained.
 
In Sections \ref{section:width} and \ref{section:constant}, we will calibrate the algorithm with extensive computer-generated numerical simulations and test it for stability and accuracy. The lessons learned from these computer simulations of events are summarized in Section \ref{section:lessons}.

Finally, in Appendix \ref{appendix} we give mathematical details about  fitting data using  the robust $\Lambda^2$ (Lorentzian) maximum likelihood estimator that we employ in our ``Sieve'' algorithm and in particular, $\Lambda^2_0$, which minimizes the rms (root mean square) widths of the parameter distributions, making them essentially the same as the rms distributions of a $\chi^2$ fit.  We also discuss fitting data with the more conventional $\chi^2$ maximum likelihood estimator. 
\section{The Adaptive Sieve Algorithm}\label{section:sieve} 
\subsection{Major assumptions}
Our major assumptions about the experimental data are: 
\begin{enumerate}
\item The experimental data can be  fitted by a model which successfully describes the data.
\item The signal data are Gaussianly distributed, with Gaussian errors.
\item That we have ``outliers'' only, so that the background consists only of points ``far away'' from the true signal. \label{outlier}
\item The noise data, {\em i.e.} the outliers, do not completely swamp the signal data. 
\end{enumerate} 
\subsection{Algorithmic steps}
We now outline our adaptive Sieve  algorithm, consisting of several steps:
\begin{enumerate}
\item{Make a robust fit (see Appendix \ref{appendix}) of {\em all} of the data (presumed outliers and all)\ by minimizing $\Lambda^2_0$, the Lorentzian squared, defined as
\be
\Lambda^2_0(\alphabold;\xbold)\equiv\sum_{i=1}^N\ln\left\{1+0.18\delchisq\right\},\label{lambda0}
\ee 
described in detail in the Appendix \ref{section:lorentz}. The $M$-dimensional parameter space of the fit is given by  $\alphabold=\{\alpha_1,\ldots,\alpha_M\}$, $\xbold=\{{x_1,\ldots,x_N}\}$ represents the abscissa of the $N$ experimental measurements $\mbox{\boldmath $y$}=\{y_1,\ldots,y_N\}$ that are  being fit and $\delchisq\equiv \left(\frac{y_i-y(x_i;\alphabold)}{\sigma_i}\right)^2$,  where $y(x_i;\alphabold)$ is the theoretical value at $x_i$ and $\sigma_i$ is the experimental error.  As discussed in Appendix {\ref{section:lorentz}}, minimizing $\Lambda^2_0$ gives  the same total $\chi^2_{\rm min}\equiv\sum_{i=1}^N \delchisq$ from \eq{lambda0} as that found in a $\chi^2$ fit,  as well as  rms widths (errors) for the parameters---for Gaussianly distributed data---that are almost the same as those found in a $\chi^2$ fit}. The quantitative measure of ``far away'' from the true signal, {\em i.e.,} point $i$ is an  outlier corresponding to Assumption (\ref{outlier}),  is the magnitude of its $\delchisq= \left(\frac{y_i-y(x_i;\alphabold)}{\sigma_i}\right)^2$. 

If $\chi^2_{\rm min}$ is satisfactory, make a conventional $\chi^2$ fit to get the errors and you are finished.   If $\chi^2_{\rm min}$ is not satisfactory, proceed to step 
 \ref{nextstep}.
\item {Using the above robust $\Lambda^2_0$ fit as the initial estimator for the theoretical curve, evaluate $\delchisq$, for each of the $N$ experimental points.}\label{nextstep}
\item A largest cut, $\delchisq_{\rm max}$, must now be selected. For example, we might start the process with $\delchisq_{\rm max}=9$. If any of the points have $\Delta \chi^2_i(x_i;\alphabold)>\delchisq_{\rm max}$, reject them---they fell through the ``Sieve''. The choice of $\delchisq_{\rm max}$ is an attempt to pick  the largest ``Sieve'' size (largest $\delchisq_{\rm max}$) that rejects all of the outliers, while minimizing the number of signal points  rejected. \label{redo}
\item Next, make a conventional $\chi^2$ fit to the sifted set---these data points are the ones that have been retained in the ``Sieve''. This  fit is used to estimate   $\chi^2_{\rm min}$.    Since the data set has been truncated by eliminating the points with $\delchisq>\delchisq_{\rm max}$, we must slightly renormalize the $\chi^2_{\rm min}$ found to take this into account, by the factor $\cal R$. This effect is discussed later in detail in Section \ref{section:lessons}.

If the renormalized $\chi^2_{\rm min}$, {\em i.e.,} ${\cal R}\times \chi^2_{\rm min}$ is acceptable---in the {\em conventional} sense, using the $\chi^2$ distribution probability function---we consider the fit of the data to the  model to be satisfactory  and proceed to the next step. If the renormalized $\chi^2_{\rm min}$ is not acceptable and $\delchisq_{\rm max}$ is not too small, we pick a smaller $\delchisq_{\rm max}$ and go back to step \ref{redo}. The smallest value of $\delchisq_{\rm max}$ that makes much sense, in our opinion, is $\delchisq_{\rm max}>2$.  After all, one of our primary assumptions is that the noise doesn't swamp the signal. If it does, then we must discard the model---we can do nothing further with this model and data set!

\item
From the  $\chi^2$ fit that was made to the ``sifted'' data in the preceding step, evaluate  the parameters $\alphabold$.
Next, evaluate the $M\times M$ covariance (squared error) matrix of the parameter space which was found in the $\chi^2$ fit. We find the new squared error matrix for the $\Lambda^2$  fit by multiplying the covariance matrix by the square of the factor $r_{\chi^2}$ (for example, as shown later in Section \ref{2}, $r_{\chi^2}\sim 1.02,1.05$, 1.11 and 1.14 for $\delchisq_{\rm max}=9$, 6, 4 and 2,  respectively ). The values of $r_{\chi^2}>1$ reflect the fact that a $\chi^2$ fit to the {\em truncated} Gaussian distribution that we obtain---after first making  a robust fit---has a rms (root mean square) width which is somewhat greater than the  rms width of the $\chi^2$ fit to the same untruncated distribution. Extensive computer simulations, summarized in Section \ref{section:lessons}, demonstrate that this {\em robust} method of error estimation yields accurate error estimates and error correlations, even in the presence of large backgrounds.
\end{enumerate}

You are now finished.  The initial robust $\Lambda^2_0$ fit has been used  to allow the phenomenologist to find a    sifted data set. The subsequent application of a $\chi^2$ fit to the {\em sifted set} gives stable estimates of the model parameters $\alphabold$, as well as a goodness-of-fit of the data to the model when $\chi^2_{\rm min}$ is renormalized for the effect of truncation due to the cut $\delchisq_{\rm max}.$   Model parameter errors are found when the covariance (squared error) matrix of the $\chi^2$ fit is multiplied by the appropriate factor $(r_{\chi^2})^2$ for the cut $\delchisq_{\rm max}$. 

It is the {\em combination} of using both  $\Lambda^2_0$ (robust) fitting  and  $\chi^2$ fitting techniques on the sifted set that gives the Sieve algorithm its power to make both a robust estimate of the parameters $\alphabold$ as well as a robust estimate of their errors, along with an estimate of the goodness-of-fit.

Using this same sifted data set, you might then try to fit  to a {\em different} theoretical model and find  $\chi^2_{\rm min}$ for this second model.  Now one can compare the probability of each model in a meaningful way, by  using the $\chi^2$ probability distribution function of the numbers of degrees of freedom for each of the models. If the second model had a very unlikely $\chi^2_{\rm min}$, it could now be eliminated.  In any event, the model maker would now  have an objective comparison of the probabilities of the two models.  
\subsection{Evaluating the Sieve algorithm}
We will give two separate types of examples  which illustrate the Sieve algorithm.  In the first type, we computer-generated   data, normally distributed about
\begin{itemize}
\item a straight line, along with random  noise to provide outliers,
\item a constant, along with random  noise to provide outliers,
\item a parabola, with background noise normally distributed about a slightly different parabola, 
\end{itemize}
the details of which are described below.  The advantage here, of course, is that we know which points are signal and which points are noise.
 
For our real world example, we took four types of experimental data for elementary particle scattering from the archives of the Particle Data Group\cite{pdg}. For all energies above 6 GeV, we took total cross sections and $\rho$-values and made a fit to these data. These were all published data points and the entire sample was used in our fit. We then made separate fits to 
\begin{itemize}
\item  $\bar pp$ and $pp$ total cross sections and $\rho$-values,
\item  $\pi^-p$ and $\pi^+p$ total cross sections $\sigma$ and $\rho$-values,
\end{itemize}
using eqns. (\ref{sigmapm}), (\ref{rhopm}) and (\ref{derivpm}) below.
 
\section{Studies using large computer-generated data sets}\label{section:computer}
Extensive  computer simulations were made using the straight line model $ y_i=1-2x_i$ and the constant model $ y_i=10$. Over 500,000 events were computer-generated, with normal distributions of 100 signal points per event, some with no noise and others with 20\% and 40\% noise added, in order to investigate the accuracy and stability of the ``Sieve'' algorithm.  The cuts $\delchi>9$, 6, 4 and 2 were investigated in detail.  
\subsection{A straight line model}\label{section:line}
An event consisted of generating  100 signal points plus either  20 or 40 background points, for a total of 120 or 140 points, depending on the background level desired.  Let RND be a random number, uniformly distributed from 0 to 1. 
Using random number generators, the first 100 points used $x_i=10\times \rnd$, where $i$ is the point number. This gives a signal randomly distributed between $x= 0$ and $x=10$. For each point $x_i$, a theoretical value $\bar y_i$ was found using $\bar y_i=1-2x_i$. Next, the value of $\sigma_i$, the ``experimental error'', {\em i.e,} the  error bar assigned to point $i$, was generated as $\sigma_i=a_i+\alpha_i\times {\rnd}$. Using these $\sigma_i$, the $y_i$'s were generated, normally distributed\cite{random} about the value of $\bar y_i$ For $i=1$ to 50, $a_i=0.2,\ \alpha_i=1.5$, and for $i=51$ to 100, $a_i=0.2,\ \alpha_i=3$. This sample of 100 points  made up the signal.

The 40 noise points, $i=101$ to 140 were generated as follows. Each point was assigned an ``experimental error'' $\sigma_i=a_i+\alpha_i\times\rnd$. The $x_i$ were generated as $x_i=d_i+\delta_i\times \rnd$. In order to provide outliers, the value of $y_i$ was {\em fixed} at $y_i=1-2x_i+f_{\rm cut}\times{\rm Sign_i}\times(b_i+\beta_i)\times\sigma_i$ and the points were then placed at this fixed value of $ y_i$ and given the ``experimental error'' $\sigma_i$.  The parameter $f_{\rm cut}$ depended only on the value of $\delchisqmax$ that was chosen, being 1.9, 2.8, 3.4 or 4, for $\delchisqmax=2$, 4, 6 or 9, respectively, and was independent of $i$. These choices of $f_{\rm cut}$ made outliers that only existed for values of $\delchisq>\delchisqmax$.

For $i=101$ to 116, $d_i=0,\ \delta_i=10,\ a_i=0.75,\ \alpha_i=0.5,\ b_i=1.0, \ \beta_i=0.6$. To make ``doubles'' at the same $x_i$ as a signal point, if $y_{i-100}>1-2x_{i-100}$ we pick  Sign$_i=+1$; otherwise Sign$_i=-1$, so that the outlier is on the same side of the reference line $1-2x_i$ as is the signal point.

For $i=117$ to 128, $d_i=0,\  \delta_i=10,\ a_i=0.5,\ \alpha_i=0.5,\ b_i=1.0, \ \beta_i=0.6$; Sign$_i$ was randomly chosen as +1 or -1. This generates outliers randomly distributed above and below the reference line, with $x_i$ randomly distributed from 0 to 10.

For $i=129$ to 140, $d_i=8,\ \delta_i=2,\ a_i=0.5,\ \alpha_i=0.5,\ b_i=1.0, \ \beta_i=0.6$; Sign$_i$ = +1. This makes points in a ``corner'' of the plot, since $x_i$ is now randomly distributed at the ``edge'' of the plot, between 8 and 10.  Further, all of this points are above the line, since Sign$_i$ is fixed at +1, giving these points a large lever arm in the fit.  

For the events generated with 20 noise points, the above recipes for background were simply halved.  An example of such an event containing 120 points, for which $\delchisqmax=6$,  is shown in Fig. \ref{noisy}a, with the 100 squares being the normally distributed data and the 20 circles being the noise data. 

After a robust fit to the entire 120 points, the sifted data set retained 100 points after the $\delchi>6$ condition was applied.  This fit had $\chi^2_{\rm min}=88.69$, with an expected $\chi^2=\nu=98$, giving $\chi^2_{\rm min}/\nu=0.905$. Using a renormalization factor  ${\cal R}=1/0.901$, we get a renormalized  $\chi^2_{\rm min}/\nu=1.01$---see Section \ref{section:lessons}  for details of the renormalization factor. After using the Sieve algorithm, by minimizing $\chi^2$ for the sifted set, we found that the best-fit straight line, $y=<a>+<b>x$, had $<a>=0.998\pm0.12$ and $<b>=-2.014\pm0.020$.   The parameter errors given above come from multiplying the errors found in a conventional $\chi^2$ fit to the sifted data by the  factor $r_{\chi^2}=1.05$---for details see Section \ref{section:lessons}.  This turns out to be a  high probability fit\cite{probability} with a  probability of 0.48 (since the renormalized $\chi^2_{\rm min}/\nu=1.01$, whereas we expect $<\chi^2/\nu>=1.0\pm0.14$).  

Figure \ref{noisy}b shows the results after the use of the Sieve procedure with $\delchimax=6$.  Of the original 120 points, all 100 of the signal points were retained (squares), while no noise points (diamonds) were retained. The solid line is the best $\chi^2$ fit,  $y=0.998-2.014x$.   

Had we applied a $\chi^2$ minimization to original 120 point data set, we would have found $\chi^2=570$, which has infinitesimal statistical probability. The straight line resulting from that fit, $y=0.925-1.98x$,  is also shown in Fig. \ref{noisy}b as the dot-dashed curve. For large x, it tends to overestimate the true values.

To investigate the stability of our procedure with respect to our choice of  $\delchi$, we reanalyzed the full data set for the cut-off, $\delchimax=4$. The evaluation of the parameters $a$ and $b$ was completely stable, essentially independent of the choice of $\delchi$.   The robustness of this procedure on this particular data set is  evident.
\subsection{Distributional widths  for the straight line model}\label{section:width}
We now generate extensive computer simulations of data sets resulting from the straight line $y_i=1-2x_i$ using the recipe of Section \ref{section:line}, with and without outliers, in order to test the Sieve algorithm.
We have generated 50,000 events with 20\% background and 50,000 events with 40\% background, for each cut $\delchimax=9$, 6, 4 and 2. We also generated 100,000 Gaussianly distributed events with no noise. 

\subsubsection{Case 1}\label{1} We  generated 100,000 Gaussianly distributed events with {\em no } noise. Let $a$ and $b$ be the intercept and slope of the straight line $y=1-2x$ and define $<a>$ as the average $a$, $<b>$ as the average $b$ found for the 100,000 straight-line events, each generated with 100 data points, using both a $\Lambda^2_0$ (robust) fit and a $\chi^2$ fit. The purpose of this exercise was to find ${r}(\Lambda^2_0)$,   the ratio of the $\Lambda^2_0$ rms  parameter width $\sigma(\Lambda^2_0)$ divided by $\Sigma$, the parameter error from the $\chi^2$ fit, {\em i.e.}
\[
r_a(\Lambda^2_0)\equiv {\sigma_a(\Lambda^2_0)\over\Sigma_a}, \quad r_b(\Lambda^2_0)\equiv {\sigma_b(\Lambda^2_0)\over\Sigma_b},
\]
as well as demonstrate that there were no biases (offsets) in  parameter determinations found in   $\Lambda^2$ and $\chi^2$ fits.

The measured offsets $1-<a_{\chi^2}>$, $1-<a_{\Lambda^2}>$, $-2-<b_{\chi^2}>$ and $-2-<b_{\Lambda^2}>$ were all numerically compatible with zero, as expected, indicating that the parameter expectations were not biased.

  
Let $\sigma$ be the rms width of a parameter distribution and $\Sigma$ the error from the $\chi^2$ covariant matrix. We found:
\begin{eqnarray*}
\sigma_a(\chi^2)&=&0.139\pm0.002\ \quad{\rm and \ }\Sigma_a=0.138\\
\sigma_b(\chi^2)&=&0.0261\pm0.003\quad {\rm and\ }\Sigma_b=0.0241, 
\end{eqnarray*}
showing that the rms widths $\sigma$ and parameter errors $\Sigma$ were the same for the $\chi^2$ fit, as expected.  
Further, the  width ratios $r$ for the $\Lambda^2_0$ fit are given by 
\begin{eqnarray*}
r_a({\Lambda^2_0})&=& 1.034\pm0.010\\
r_b(\Lambda^2_0)&=& 1.029\pm0.011,
\end{eqnarray*}
demonstrating that:
\begin{itemize}
\item the $r$'s of the $\Lambda^2_0$ are almost as good as that of the $\chi^2$ distribution, $r(\chi^2)=1$.
\item the  ratios   of the rms $\Lambda^2$ width to the rms $\chi^2$ width for both  parameters $a$ and $b$ are the {\em same}, {\em i.e.,} we can now simply write 
\be
r_{\Lambda^2}={\sigma_{\Lambda^2}\over{\Sigma}}\sim 1.03.
\ee
\end{itemize}
 Finally, we find that $1-<\chi^2/\nu>=0.00034\pm0.00044$, which  is approximately  zero, as expected.
\subsubsection{Case 2}\label{2}
For Case 2, we investigate data generated with 20\% and 40\% noise that  have been subjected to  the  adaptive Sieve algorithm, {\em i.e.} the sifted data  after cuts of $\delchisqmax=9$, 6, 4 and 2. We investigated this truncated sample to measure  possible biases and  to obtain numerical values for $r$'s.

We generated 50,000 events, each with 100 points normally distributed and with either 20 or 40 outliers, for each cut. A robust fit was made to the entire sample (either 120 or 140 points) and we 
sifted the data, rejecting all points with either $\delchisq>9,$ 6, 4 and 2, according to how the data were generated. A conventional $\chi^2$ analysis was then made to the sifted data. The results are summarized in Table \ref{table:rchisq}. As before, we found that the widths from the $\chi^2$ fit were slightly smaller than the widths from a robust fit, so we adopted only the results for the $\chi^2$ fit.

There were negligible offsets $1-<a>$ and $-2-<b>$, being $\sim 1$ to $ 5\%$ of the relevant rms widths, $\sigma_a$ and $\sigma_b$, for both the robust and $\chi^2$ fits. 

In any individual $\chi^2$ fit to the $j$th data set, one measures $a_j,b_i,\Sigma_{a_j},\Sigma_{b_j}$ and $(\chi^2_{\rm min}/\nu)_j$. Thus, we  characterize all of our computer simulations in terms of these 7 observables. 

We again  find  that the $r_{\chi^2}$ values---defined as $\sigma/\Sigma$---are the same, whether we are measuring $a$ or $b$. They are given by 
$r_{\chi^2}={\sigma}/{\Sigma}=1.034$, 1.054, 1.098 and 1.162  
for the cuts $\delchisqmax=9,$ 6, 4 and 2, respectively\cite{subtle}. Further, they are the same for 20\% noise and 40\% noise, since the cuts rejected all of the noise points.  In addition, the $r$ values  were found to be the same as the $r$ values for the  case of truncated pure signal, using  the same $\delchisqmax$ cuts.  The signal retained was 99.7, 98.57, 95.5 and 84.3 \% for the cuts  $\delchisqmax =9,$ 6, 4 and 2, respectively---see Section \ref{section:lessons} and \eq{SF} for theoretical values of the amount of signal retained.

We experimentally determine the rms widths $\sigma$ (the errors of the parameter) by multiplying the $r$ value,  a known  quantity {\em independent} of the particular event, by the appropriate $\Sigma$ which is measured for that event, {\em i.e.,}
\begin{eqnarray*}
{\sigma_a}&=&\Sigma_a\times r_{\chi^2}\\
{\sigma_b}&=&\Sigma_b\times r_{\chi^2}.
\end{eqnarray*}
The  rms widths are now determined for {\em any} particular data set by multiplying the known factors $r_{\chi^2}$ by the appropriate $\Sigma$ found (measured) from the covariant matrix of the $\chi^2$ fit of that data set.

Also shown in Table \ref{table:rchisq} are the values of $\chi^2_{\rm min}/\nu$ found for the various cuts. We will compare these results later with those for the constant case, in   Section \ref{section:constant}

We again see that a sensible approach for data analysis--even where there are large backgrounds of $\sim 40\%$---is to use the parameter estimates for $a$ and $b$ from the truncated $\chi^2$\  fit and assign  their errors as 
\begin{eqnarray}
\sigma_a&=r_{\chi^2}\Sigma_a \nonumber\\
\sigma_b&=r_{\chi^2}\Sigma_b,
\end{eqnarray}
where $r_{\chi^2}$ is a function of the $\delchimax$ cut utilized.
Before estimating the goodness-of-fit, we must renormalize the observed $\chi^2_{\rm min}/\nu$ by the appropriate numerical factor for the $\delchimax$ cut used.

This strategy of using  an adaptive $\delchisqmax$ cut minimizes the error assignments,  guarantees  robust fit parameters with no significant bias and also returns a goodness-of-fit estimate.

\subsection{The constant model, $y_i=10$}\label{section:constant}
For this case, we investigate a different theoretical model ($y_i=10$) with a different background distribution, to measure the values of $r_{\chi^2}$ and $<\chi^2_{\rm min}/\nu>$. 

An event consisted of generating  100 signal points plus either  20 or 40 background points, for a total of 120 or 140 points, depending on the background level desired.  Again, let RND be a random number, uniformly distributed from 0 to 1. 
Using random number generators, for the  first 100 points $i$,  a theoretical value $ \bar y_i=10 $ was chosen. Next, the value of $\sigma_i$, the ``experimental error'', {\em i.e,} the  error bar assigned to point $i$, was generated as $\sigma_i=a_i+\alpha_i\times {\rnd}$. Using these  $\sigma_i$, the $y_i$'s were generated, normally distributed\cite{random} about the value of $\bar y_i=10$ . For $i=1$ to 50, $a_i=0.2,\ \alpha_i=1.5$, and for $i=51$ to 100, $a_i=0.2,\ \alpha_i=3$. This sample of 100 points  made up the signal.

The 40 noise points, $i=101$ to 140 were generated as follows. Each point was assigned an ``experimental error'' $\sigma=a_i+\alpha_i\times\rnd$. In order to provide outliers, the value of $y_i$ was fixed at $y_i=10+f_{\rm cut}\times{\rm sign_i}\times(b_i+\beta_i)\times\sigma_i$ and the points were then placed at this fixed value of $ y_i$ and given the ``experimental error'' $\sigma_i$.  The parameter $f_{\rm cut}$ depended only on the value of $\delchisqmax$ that was chosen, being 1.9, 2.8, 3.4 or 4, for $\delchisqmax=2$, 4, 6 or 9, respectively, and was independent of $i$.

For $i=101$ to 116, $a_i=0.75,\ \alpha_i=0.5,\ b_i=1.0, \ \beta_i=0.6$; Sign$_i$ was randomly chosen at +1 or -1. 

For $i=117$ to 128, $a_i=0.5,\ \alpha_i=0.5,\ b_i=1.0, \ \beta_i=0.6$; This generates outliers randomly distributed above and below the reference line, with $x_i$ randomly distributed from 0 to 10.

For $i=129$ to 140, $a_i=0.5,\ \alpha_i=0.5,\ b_i=1.0, \ \beta_i=0.6$; Sign$_i$ = +1. This forces 12 points to be greater than 10, since Sign$_i$ is fixed at +1.  For the events generated with 20 noise points, the above recipes for background were simply halved. 

Two examples of events with 40 background points are shown in Figures \ref{constant140_4}a and \ref{constant140}a, with the 100 squares being the normally distributed data and the 40 circles being the noise data. 

In Fig. \ref{constant140_4}b we show the results after using the cut $\delchisqmax=4$.  No noise points (diamonds)  were retained, and 98 signal  points (circles) are shown. The best fit, $y=9.98\pm0.074$, is the solid line, whereas the dashed-dot curve is the fit to all 140 points.  The observed $\chi^2_{\rm min}/\nu=0.84$ yields a renormalized value ${\cal R}\times \chi^2_{\rm min}/\nu=1.09$, in good agreement with the expected value $\chi^2_{\rm min}/\nu=1\pm0.14$. If we had fit to the entire 140 points, we would find $\chi^2_{\rm min}/\nu=4.39$, with the fit being the dashed-dot curve.  

In Fig. \ref{constant140}b we show the results after using the cut $\delchisqmax=9$.  No noise points (diamonds)  were retained, and 98 signal  points (circles) are shown. The best fit, $y=10.05\pm0.074$, is the solid line, whereas the dashed-dot curve is the fit to all 140 points.  The observed $\chi^2_{\rm min}/\nu=1.08$ yields a renormalized value ${\cal R}\times \chi^2_{\rm min}/\nu=1.11$, in good agreement with the expected value $\chi^2_{\rm min}/\nu=1\pm0.14$. If we had fit to the entire 140 points, we would find $\chi^2_{\rm min}/\nu=8.10$, with the fit being the dashed-dot curve.  The details of the renormalization of $\chi^2_{\rm min}/\nu$ and the assignment of the errors are given in Section \ref{section:lessons}

We computer-generated a total of 500,000 events,  50,000 events with  20\% noise and an additional 50,000 events with 40\% noise, for each of the cuts $\delchi>9$, 6, 4 and 2, and 100,000 events with no noise. 

For the sample with no cut and no noise, we found $r_{\Lambda^2_0}=1.03\pm0.02$,  equal to the value  $r_{\Lambda^2_0}=1.03$ that was found for the straight line case.

Again, we found that our results for $r_{\chi^2}$ were independent of background, as well as model, and only  depended on the cut.  We also found that the biases (offsets) for the constant case, $(10-<a_{\chi2}>)$, although non-zero for the noise cases,  were small in comparison to   $\sigma$, the rms width.

The results for cuts $\delchimax=9$, 6, 4 and 2  are detailed in Table \ref{table:rchisq}.  
We see in Table \ref{table:rchisq}, compared with the straight line results of Section \ref{2}, that the $r_{\chi^2}$ values for the constant case are essentially identical, as expected. Further, we find the same results for the values of $\chi^2_{\rm min}/\nu$ as a function of the cut $\delchisqmax$.

\subsection {Lessons learned from computer studies  of  a straight line model and  a constant model}\label{section:lessons}
\begin{itemize}
\item
As found in Sections \ref{2} and \ref{section:constant} and  detailed in Table \ref{table:rchisq}, we have universal values of $r_{\chi^2}$ and $<\chi^2_{\rm min}>/\nu$,  as a function of the cut $\delchisqmax$, independent of both background and model.
\item A sensible conservative approach for large backgrounds  (less than or the order 40\%) is to use the parameter estimates from the $\chi^2$ fit to the sifted data and assign  the parameter errors to the fitted robust parameters as 
\begin{eqnarray}
\sigma(\chi^2)&=&r_{\chi^2}\times\Sigma,\nonumber
\end{eqnarray}
where $r_{\chi^2}$ is a function of the cut $\delchisqmax$, given by  the average of the straight line and constant cases of Table \ref{table:rchisq}.
This strategy gives us a minimum parameter error, with only very small  biases to the parameter estimates.
\item  We must then renormalize the value found for $\chi^2_{\rm min}/\nu$ by the appropriate averaged value of $<\chi^2_{\rm min}>/\nu$  for the straight line and  constant case, again as a function of the cut $\delchisqmax$. 
\item Let us define $\Delta$ as the $\delchimax$ cut and $\cal R$ as the renormalization factor that multiplies $\chi^2_{\rm min}/\nu$.

We find from inspection of Cases 1 to 2 for the straight line and  of  Section \ref{section:constant} for the case of the constant fit that  a best fit parameterization of $r_{\chi^2}$, valid for $ \Delta\ge2$\ is given by 
\be
r_{\chi^2}=1+0.246e^{-0.263\Delta}.\label{rho}
\ee
 
We note that ${\cal R}^{-1}$, for large $\nu$,  is given analytically by
\begin{eqnarray}
{\cal R}^{-1}&\equiv&{\int^{+\sqrt{\Delta}}_{-\sqrt\Delta} x^2e^{-x^2/2}\,dx}/{\int^{+\sqrt{\Delta}}_{-\sqrt\Delta} e^{-x^2/2}\,dx}\nonumber\\
&=&1-\frac{2}{\sqrt{\pi}}\frac{e^{-\Delta/2}}{{\rm erf}(\sqrt{\Delta/2})}.\label{Rminus1}
\end{eqnarray}
Graphical representations of $r_{\chi^2}$ and ${\cal R}^{-1}$  are shown in Figures  \ref{renorm}a and \ref{renorm}b, respectively. Some numerical values are given in Table \ref{table:rchisq} and are compared to the computer-generated values found numerically for the straight line and constant cases. The agreement is excellent.
\item Let us define $\sigma_0$ as the rms parameter width  that we would have had for a $\chi^2$ fit to the uncut sample, where the sample had  had no background, and define $\Sigma_0$ the error found from the covariant matrix. They are, of course, equal to each other, as well as being the smallest error possible. We note that the ratio $\sigma/\sigma_0=r_{\chi^2}\times\Sigma/\Sigma_0$. This ratio is a function of the cut $\Delta$  through both $r_{\chi^2}$ {\em and} $\Sigma$, since for a truncated distribution, $\Sigma/\Sigma_0$ depends inversely on the square root  of the fraction of signal points that survive the cut $\Delta$. In particular, the survival fraction $S. F.$ is given by
\be
S.F.=\int^{+\sqrt{\Delta}}_{-\sqrt\Delta}\frac{1}{\sqrt{2\pi}} e^{-x^2/2}\,dx= {\rm erf}(\sqrt{\Delta/2})\label{SF}
\ee
and is  99.73, 98.57, 95.45 and 84.27 \% for the cuts  $\Delta =9,$ 6, 4 and 2, respectively. The survival fraction $S.F.$ is shown in Table \ref{table:rchisq} as a function of the cut $\delchimax$, as well as is the ratio $\sigma/\sigma_0$.  We note that the true cost of truncating a Gaussian distribution, {\em i.e.,} the enlargement of the error due to truncation,  is not $r_{\chi^2}$, but rather $r_{\chi^2}/\sqrt{S.F.}$, which  ranges from $\sim 1.02$ to 1.25 when the cut $\delchimax$ goes from 9 to 2. This rapid loss of accuracy is why the errors become intolerable for cuts $\delchimax$ smaller than 2.  

\end{itemize}
\subsection {Fitting strategy}\label{section:strategy}

We find that an effective  strategy for eliminating noise and making robust parameter estimates, together  with {\em robust}   error assignments, is:
\begin{enumerate}
\item Make an initial $\Lambda^2_0$ fit to the entire data sample. If $\chi^2_{\rm min}/\nu$ is satisfactory, then make a standard $\chi^2$ fit to the data and you are finished. If not, then proceed to the next step.
\item Pick a large value of $\delchisq_{\rm max}$, {\em e.g.,} $\delchisq_{\rm max}=9.$ 
\item Obtain a sifted sample by throwing away all points with $\delchisq>\delchisq_{\rm max}$.\label{sifted} 
\item  Make  a conventional $\chi^2$  fit to the sifted sample. For your choice of $\delchisq_{\rm max}$, find ${\cal R}^{-1}$ from \eq{Rminus1}. If the renormalized value ${\cal R}\times \chi^2_{\rm min}/\nu$ is sufficiently near 1, {\em i.e.,} the goodness-of-fit is satisfactory, then go to the next step. If, on the  other hand,
${\cal R}\times \chi^2_{\rm min}/\nu$ is too large, pick a smaller value of $\delchisq_{\rm max}$ and go to step \ref{sifted} (for example, if you had used a cut of 9, now pick $\delchisq_{\rm max}=6$ and start again).  Finally, if you reach $\delchisq_{\rm max}=2$ and you still don't have success, quit---the background has penetrated too much into the signal for the ``Sieve'' algorithm to work properly.  
\item  a) Use the parameter estimates found from the $\delchisq_{\rm max}$ fit in the previous step.\\ b) Find a  new squared error matrix  by multiplying the  covariant matrix $C$ found in the $\chi^2$ fit by $(r_{\chi^2})^2$. Use the value of $r_{\chi^2}$ found in \eq{rho} for the chosen value of the cut  $\delchisq_{\rm max}$  to obtain a robust error estimate essentially independent of background distribution.\\\\
 You are now finished. You have made a robust determination of the parameters, their errors and the goodness-of-fit.  
\end{enumerate}
The renormalization factors $\cal R$ are only used in  estimating the value of the goodness-of-fit, where small changes in this value are not very important.   Indeed, it hardly matters if the estimated renormalized $\chi^2/\nu$ is  between 1.00 and 1.01---the possible variation of the expected renormalized $\chi^2/\nu$ due to the two different background distributions.  After all,  it is a subjective judgment call on the part of the phenomenologist as to whether the goodness-of-fit is satisfactory.  For large $\nu$, only when $\chi^2/\nu$ starts approaching 1.5  does one really begin to start worrying  about the model. For $\nu\sim100$, the error expected in $\chi^2/\nu$ is $\sim 0.14$, so uncertainties in the renormalized $\chi^2/\nu$ of the order of several percent play no critical role. The accuracy of the renormalized values is perfectly adequate for the  purpose of judging whether  to keep or discard a model.  

In summary, extensive computer simulations for  {\em sifted data sets}  show that by  combining the  $\chi^2$ parameter determinations  with the corrected covariance matrix from the $\chi^2$ fit, we obtain also a ``robust'' estimate of the errors, basically independent of both the background distribution and the model. Further, the renormalized $\chi^2_{\rm min}/\nu$ is a good predictor of the goodness-of-fit.  Having to make  a $\Lambda^2_0$ fit to sift the data and then a $\chi^2$ fit to the sifted data is a small computing cost to pay  compared to the ability to make accurate predictions. Clearly, if the data are not badly contaminated with outliers, {\em e.g.,} if a $\delchisq_{\rm max}=6$ fit is satisfactory, the additional penalty paid is that the errors are enlarged by a factor of $\sim1.06$ (see Table \ref{table:rchisq}), which is not unreasonable to rescue a data set. Finally, if you are not happy about the error determinations, you can use the parameter estimates you have found to make Monte Carlo simulations of your model\cite{nr}.  By repeating a $\Lambda_0^2$ fit to the simulated distributions and then sifting them to the same value of $\delchisq_{\rm max}$ as was used in the initial determination of the parameters, and finally, by making a $\chi^2$ fit to the simulated sifted set  you can make an error determination based on the spread in the parameters found from the simulated data sets.

\subsection{The parabola}\label{section:parabola}
As a final example of computer-generated data, we generated one noisy data set using  a parabolic model.
A total of 135 points were generated by computer. 
Using random number generators, the first 50 points generated picked $x_i$'s  distributed randomly\cite{random} from 0 to 10.  For each point $x_i$, a theoretical value $\bar y_i$ was found using $\bar y_i=1+2x_i+0.5x_i^2$. Next, the value of $\sigma_i$, the ``experimental error'', {\em i.e,} the  error bar assigned to point $i$, was generated randomly on the interval 0.2 to 2.7. The $y_i$'s were then generated, normally distributed\cite{random} about the value of $\bar y_i$ using the $\sigma_i$ that had been previously found. The next 50 points were chosen in the same manner, except that these $\sigma_i$ were randomly distributed between 0.2 and 5.2. This sample of 100 points  made up the signal.

The 35 noise points were generated around a ``nearby'' parabola, given by $ \bar y_i=12+2x_i+0.2x_i^2$.  The first 15 points had their $x_i$ again randomly generated in the interval 0 to 10. The error bars assigned to each point were randomly distributed in the interval 0.2 to 5.2.  To provide the outliers, the value of the theoretical $\bar y_i$ was found using a new parabola $\bar y_i=12+2x_i+0.2x_i^2$. These points were then normally distributed using $\sigma_i$'s  uniformly distributed in the interval 0.8 to 20.8.  The next 20 were generated in the same fashion, except that the error bars were uniformly distributed in the interval 0.2 to 8.2 and the $y_i$ values normally distributed with $\sigma_i$'s in the interval 1.6 to 65.6.
 In this case, we not only made ``outliers'', but also contaminated the sample with substantial ``inliers'', since we used a ``nearby parabola'' to generate the background data.  Of course, this violates our Assumption \ref{outlier} that we only have outliers, but gives us a feeling of what happens if substantial amounts of ``inliers'' are also present.

 The resulting distribution of 135 points is shown in Fig. \ref{noisyparabola}a, with the 100 squares being the normally distributed data and the 35 circles being the noise data.  

The sifted data set, shown in Fig. \ref{noisyparabola}b,  retained 113 points after the $\delchimax=6$ condition was applied to the original 135 points.  At that point, we made both a conventional $\chi^2$ fit to the sifted data set in order to evaluate  the parameters, their errors and the goodness of fit. The $\chi^2$ fit to the sifted data had $\chi^2_{\rm min}=123.6$, with $\nu=110$, giving $\chi^2_{\rm min}/\nu=1.12$. Renormalizing using $\cal R$ found from \eq{Rminus1} }, we get the corrected ${\cal R}\times\chi^2_{\rm min}/\nu=1.24$, whereas we expect $1\pm0.13$.  This is a reasonable fit\cite{probability} with a probability of $\sim 0.06$.  After using the Sieve algorithm, by minimizing $\chi^2$, we found that the best-fit parabola, $y=c_0+c_1x+c_2x^2$, had $c_0=1.18\pm0.23$ and $c_1=2.05\pm0.05$ and $c_2=0.489\pm0.005$, where the errors have been renormalized by the factor $r_{\chi^2}=1.05$ found from \eq{rho}. 

Figure \ref{noisyparabola}b shows the results of using the Sieve procedure with the cut $\delchisqmax=6$.  Of the original 135 points, all 100 of the signal points were retained (squares). There were 13  noise points (circles) also retained, all very  close to the fitted straight line. These points are the ``inliers'' that resulted from the background generation using the ``nearby parabola'', violating our primary assumption that there are only ``outliers'' as background. Thus, it is of great interest to see how well the Sieve procedure worked. 

Had we applied a $\chi^2$ minimization to original 130 point data set, we would have found $\chi^2_{\rm min}/\nu=19.93$, which clearly has infinitesimal statistical probability. The parabola  resulting from this $\chi^2$  fit is also shown in Fig. \ref{noisyparabola}b. It clearly misses many of the  data points in the sifted set. 

When we fitted the parabola to  {\em only} the 100 signal points, with no noise included, we got the parameters: $c_0=0.97\pm0.21,\ c_1=2.13\pm0.05$ and $c_2=0.480\pm0.005$, using a conventional $\chi^2$ fit. These parameters, within errors the {\em same} as those found using the ``Sieve'' algorithm, give a curve that is essentially indistinguishable from the solid line in Fig. \ref{noisyparabola}b obtained using the Sieve algorithm. We note that   even when the background produces some ``inliers'', {\em i.e.,} the cut $\delchimax$ does {\em not} remove all of the background, the Sieve procedure is still very useful.  

Finally, our procedure  was completely stable for reasonable choices of $\delchi$, giving essentially  the same answer for $\delchi >4$, 6 or 9. Thus, even in the presence of $\sim 13 \%$ ``inliers'', the answer after using the ``Sieve'' was reasonable. The parameter values are relatively unaffected, as are the errors.  The main concern is the higher corrected $\chi^2_{\rm min}/\nu$ that is due to the background points that are close to the true signal and thus can not be ``Sieved'' out. However, this only affects the goodness-of-fit estimate, making $\chi^2_{\rm min}/\nu$ somewhat larger. In the end, the conclusion as to whether to accept the model or reject it on the basis of the goodness-of-fit estimate  is  a subjective judgment of the phenomenologist.  Many  models have been accepted when the $\chi^2$ probability has been as low as a few tenths of a percent.  

\subsection{Real World data}
We will illustrate the Sieve algorithm by
simultaneously fitting  all of the published experimental data above $\sqrt s>6$ GeV  for both the total cross sections $\sigma$ and $\rho$ values for $\bar pp$ and $pp$ scattering, as well as for $\pi^-p$ and $\pi^+p$ scattering.  The $\rho$ value is the ratio of the real to the imaginary forward scattering amplitude and $\sqrt s$ is the cms energy $E_{\rm cms}$.  The data sets used have been taken from the Web site of the Particle Data Group\cite{pdg} and have not been modified. They provide the energy ($x_i$), the measurement value ($y_i$) and the experimental error($\sigma_i$), assumed to be a standard deviation, for each experimental point.

Testing the hypothesis that the cross sections rise asymptotically as $\ln^2s$, as $s\rightarrow\infty$, the four functions $\sigma^\pm$ and $\rho^\pm$ that we will {\em simultaneously} fit for $\sqrt s>6$ GeV are:
 \begin{eqnarray}
\sigma^\pm&=&c_0+c_1\ln\left(\frac{\nu}{m}\right)+c_2\ln^2\left(\frac{\nu}{m}\right)+\beta_{\cal P'}\left(\frac{\nu}{m}\right)^{\mu -1}\pm\  \delta\left({\nu\over m}\right)^{\alpha -1},\label{sigmapm}\\
\rho^\pm&=&{1\over\sigma^\pm}\left\{\frac{\pi}{2}c_1+c_2\pi \ln\left(\frac{\nu}{m}\right)-\beta_{\cal P'}\cot({\pi\mu\over 2})\left(\frac{\nu}{m}\right)^{\mu -1}+\frac{4\pi}{\nu}f_+(0)
\pm \delta\tan({\pi\alpha\over 2})\left({\nu\over m}\right)^{\alpha -1} \right\}\label{rhopm},\\
\frac{d\sigma^{\pm}}{d(\x}&=&c_1\left\{\frac{1}{(\x}\right\} +c_2\left\{ \frac{2\ln(\x}{(\x}\right\}+\beta_{\cal P'}\left\{(\mu-1)(\x^{\mu-2}\right\}\nonumber\\
&&\ \ \ \ \ \ \ \ \ \ \ \ \ \ \ \ \ \ \ \ \  \pm \ \delta\left\{(\alpha -1)(\x^{\alpha - 2}\right\},\label{derivpm}
\end{eqnarray}
where the upper sign is for $pp$ ($\pi^+p$) and the lower sign is for $\bar p p$ ($\pi^-p$) scattering\cite{physics}.
Here, $\nu$ is the laboratory energy of the projectile particle and $m$  is the proton (pion) mass. The exponents $\mu$ and $\alpha$ are real, as are the 6 constants $c_0,\ c_1,\ c_2,\ \beta_{\cal P'},\ \delta$ and the dispersion relation subtraction constant $f_{+}(0)$.  We  set $\mu= 0.5$, appropriate for a Regge-descending trajectory, leaving us 7 parameters. We then require the fit to be anchored by the experimental values of $\sigma_{\bar pp}$ and $\sigma_{pp}$ ($\sigma_{\pi^-p}$ and $\sigma_{\pi^+p}$), as well as their slopes, $\frac{d\sigma^{\pm}}{d(\x}$, at $\sqrt s =4$ GeV for nucleon scattering and $\sqrt s =2.6$ GeV  for pion scattering.  This in turn  imposes 4 conditions on the above equations and we thus have three free parameters to fit: $c_1,\ c_2$ and $f_+(0)$. 

\subsubsection{$\bar pp$ and $pp$ scattering}\label{section:pp}
The raw experimental data for $\bar pp$ and $pp$ scattering  that are shown in Figures \ref{sigpdg} and \ref{rhopdg} were taken from the Particle Data Group\cite{pdg}. Figure \ref{sigpdg} shows the $\sigma_{\bar pp}$ and $\sigma_{pp}$ data for ${\rm E}_{\rm cms}>6$ GeV, whereas Fig. \ref{rhopdg} shows all of the experimental $\rho_{\bar pp}$ and $\rho_{pp}$  data for ${\rm E}_{\rm cms}>6$ GeV. There are a total of 218 points in these 4 data sets.  We fit these 4 data sets {\em simultaneously} using \eq{sigmapm}, \eq{rhopm} and \eq{derivpm}.  Before we applied the Sieve, we obtained  $\chi^2_{\rm min}=1185.6$, whereas we expected 215.  Clearly, either the model doesn't work or there are a substantial number of outliers giving very large $\delchi$ contributions. The Sieve technique shows the latter to be the case.

We now study the effectiveness and stability of the Sieve. Table \ref{ppfit} contains the fitted results for $\bar pp$ and $pp$ scattering using 3 different choices of the cut-off, $\delchimax=4$, 6 and 9.  For each $\delchimax$ cut it tabulates: 
\begin{itemize}
\item the fitted parameters from the $\chi^2$ fit together with the errors found in the $\chi^2$ fit,
\item the total $\chi^2_{\rm min}$,
\item $\nu$,  the number of degrees of freedom (d.f.) after the data have been sifted by the indicated $\delchi$ cut-off.
\end{itemize}
To get  robust errors, the errors quoted in Table \ref{ppfit} for  for each parameter should be multiplied by the common factor $r_{\chi^2}$=1.05, from \eq{rho}, using the cut $\Delta=6$. 

We note that  for $\delchimax=6$, the number of retained data points is 193, whereas we started with 218, giving a background of $\sim13\%$. We have rejected 25 outlier points (5 $\sigma_{pp}$, 5 $\sigma_{\bar pp}$, 15 $\rho_{pp}$ and no $\rho_{\bar pp}$ points)   with $\chi^2_{\rm min}$ changing from 1185.6 to 182.8. We find  $\chi^2_{\rm min}/\nu=0.96$, which when renormalized using \eq{Rminus1} for $\Delta=6$ becomes ${\cal R}\times \chi^2_{\rm min}/\nu= 1.067$,  a very likely value  with a probability\cite{probability} of $\approx 0.25$. 

Obviously, we have cleaned up the sample---we have rejected 25 datum points which had an average $\Delta\chi^2_i\sim 40$! We have demonstrated that:
(1) the goodness-of-fit of the model is excellent, and (2) we had very large $\delchi$ contributions from the outliers that we were able to Sieve out. These outliers, in addition to giving a huge $\chi^2_{\rm min}/\nu$,  severely distort the parameters found in a conventional $\chi^2$ minimization, whereas they were easily handled by a robust fit which minimized $\Lambda^2_0$, followed by a $\chi^2$ fit to the sifted data.

Inspection of  Table \ref{ppfit} shows that the parameter values $c_1$, $c_2$ and $f_+(0)$ effectively do not depend on $\delchimax$, our  cut-off choice, having only very small changes compared to the predicted parameter errors.   

A further indication of the stability of the Sieve is illustrated in Table \ref{predicted}.  As a function of $\sqrt s$, we have tabulated:
\begin{itemize}
\item the predicted total cross sections and $\rho$-values for $\bar pp$ and $pp$
\item the errors in their predictions generated by the errors in the fit parameters $c_1,\ c_2$ and $f_+(0)$,
\end{itemize}
 for two different cut-off values,  $\delchimax=4$ and 6. The predicted cross sections and $\rho$-values for the two values of $\delchimax$  are virtually indistinguishable, giving us strong confidence in the Sieve technique when used with four different types of real-world experimental data.

\begin{figure}[h,t,b,p] 
\begin{center}
\mbox{\epsfig{file=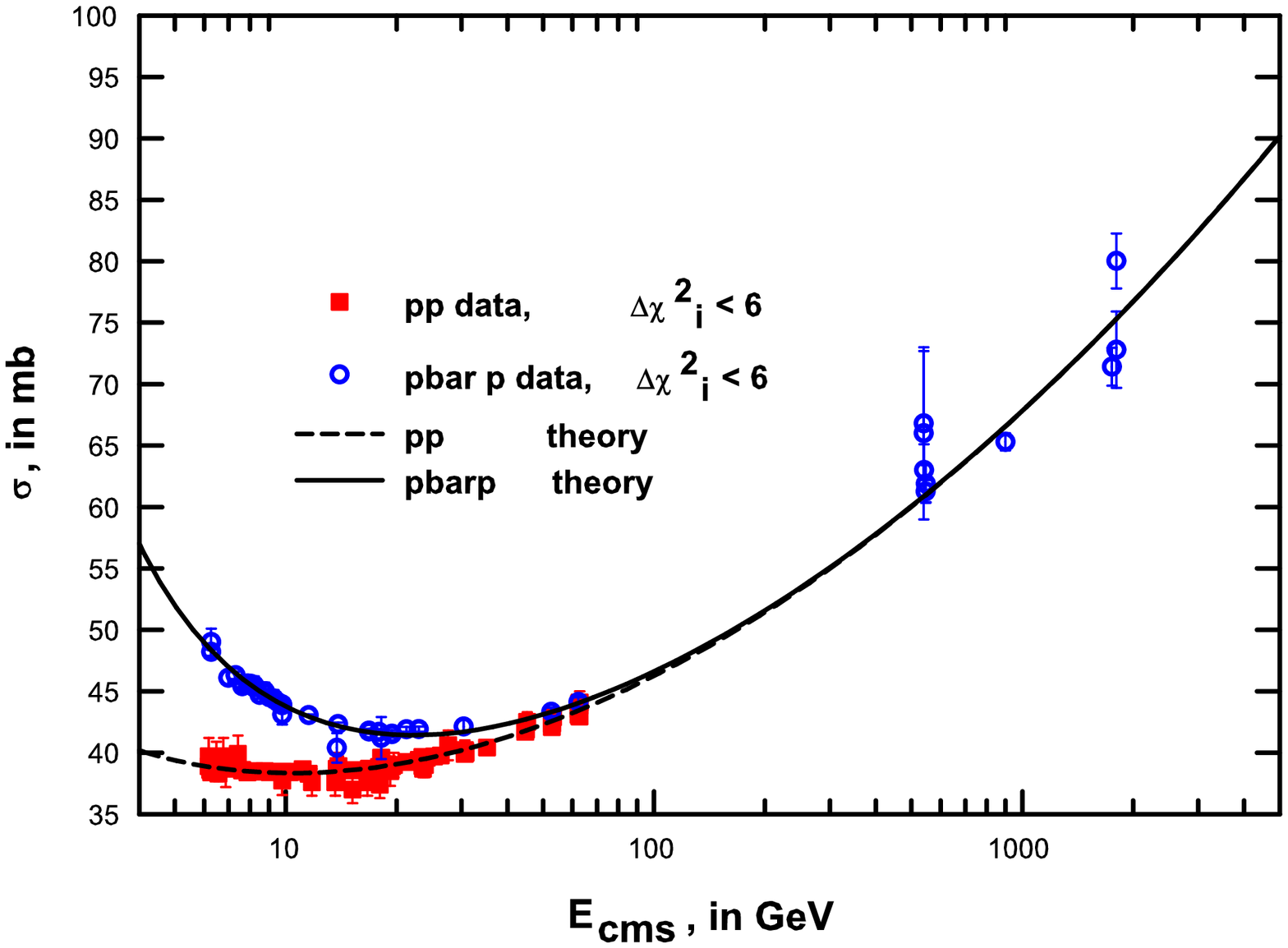,width=5in%
,bbllx=0pt,bblly=0pt,bburx=560pt,bbury=410pt,clip=%
}}
\end{center}
\caption[]{ \footnotesize
The data points shown are the result of screening all of the points of Fig. \ref{sigpdg} for those cross section points with $\Delta \chi^2_i<6.$ The open circles are $\sigma_{\bar pp}$ and the squares are $\sigma_{ pp}$. The solid line is the theoretical fit to $\sigma_{\bar pp}$ and the dashed line is the theoretical fit to $\sigma_{pp}$. 
  }
\label{sighi}
\end{figure}

\begin{figure}[h,t,b,p] 
\begin{center}
\mbox{\epsfig{file=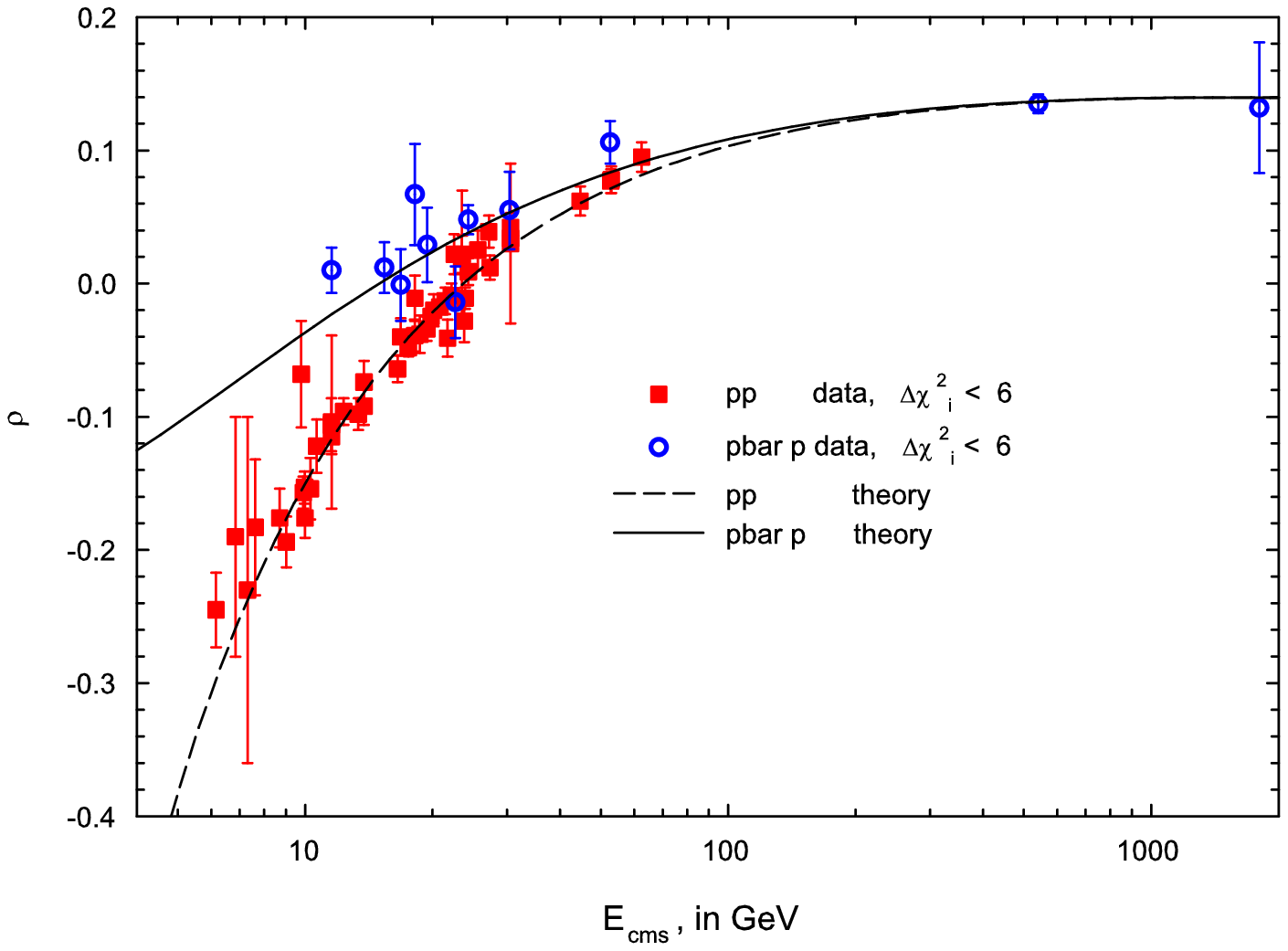,width=5in%
,bbllx=0pt,bblly=0pt,bburx=418pt,bbury=310pt,clip=%
}}
\end{center}
\caption[]{ \footnotesize
The data points shown are the result of screening all of the points in Fig. \ref{rhopdg} for those $\rho$-value points with $\Delta \chi^2_i<6.$  The open circles are $\rho_{\bar pp}$ and the squares are $\rho_{ pp}$. The solid line is the theoretical fit to $\rho_{\bar pp}$ and the dashed line is the theoretical fit to $\rho_{pp}$. 
  }
\label{rhosmall}
\end{figure}

The results of applying the Sieve algorithm to the 4 data sets, along with the fitted curves, are graphically shown in Fig. \ref{sighi} for $\sigma_{\bar pp}$ and $\sigma_{pp}$ and in Fig. \ref{rhosmall}  for $\rho_{\bar pp}$ and $\rho_{pp}$.  The total number of data points shown in Fig. \ref{sighi} and in Fig. \ref{rhosmall} is 193, whereas we started with 218 points. The fits shown are in excellent agreement with the 193 data points.

As a final test, we tried fitting another model which had its cross section energy dependence asymptotically rising as $\ln s$. This is the equivalent of setting the parameter $c_2=0$, leaving us two free parameters to fit, $c_1$ and $f_+(0)$.  Using the same sifted data set which had given $\chi^2_{\rm min}=182.8$ for the $\ln^2 s$ model we now obtained $\chi^2_{\rm min}=1185.6$ for only one more degree of freedom, clearly indicating that the $\ln s$  model was a very bad fit and could be excluded, whereas the $\ln ^2s$ model gave a very good fit to the same data subset. 
\subsubsection{$\pi^-p$ and $\pi^+p$ scattering}\label{section:pip}

The raw experimental data for $\pi^-p$ and $\pi^+p$ scattering  shown in Figures \ref{pipdg} and \ref{pirhopdg} were taken from the Particle Data Group\cite{pdg}. For ${\rm E}_{\rm cms}>6$ GeV, Figure \ref{pipdg} shows the $\sigma_{\pi^-p}$ and $\sigma_{\pi^+p}$ data and Fig. \ref{pirhopdg} shows the $\rho_{\pi^-p}$ and $\rho_{\pi+p}$  data. There are a total of 155 points in these 4 data sets.  Before we applied the Sieve algorithm, we obtained  $\chi^2=527.8$, whereas we expected 152, leading us to conclude that  either the model doesn't work or there are a substantial number of outliers giving very large $\delchi$ contributions. Once again, the Sieve technique shows the latter to be the case.

Table \ref{pipfit} contains the fitted results for $\pi^-p$ and $\pi^+p$ scattering using 3 different choices of the cut-off, $\delchimax=4$, 6 and 9.
 For each $\delchimax$ it tabulates: 
\begin{itemize}
\item the fitted parameters from the $\chi^2$ fit together with the errors found in the $\chi^2$ fit,
\item the total $\chi^2_{\rm min}$,
\item $\nu$,  the number of degrees of freedom (d.f.) after the data have been sifted by the indicated $\delchimax$ cut-off.
\end{itemize}
To get  robust errors, the errors quoted in Table \ref{pipfit} for $\delchisqmax=6$ for each parameter should be multiplied by the common factor $r_{\chi^2}$=1.05 of \eq{rho} for the cut  $\Delta=6$. 

For $\delchimax=6$,  the number of retained data points is 130, whereas we started with 155, a  background of $\sim 19\%$.  We have rejected 25 outlier points (2 $\sigma_{\pi^+p}$, 19 $\sigma_{\pi^-p}$, 4 $\rho_{\pi^+p}$ and no $\rho_{\pi^-p}$ points)   with $\chi^2_{\rm min}$ changing from 527.8 to 148.1. We find  $\chi^2_{\rm min}/\nu=1.166$, which when renormalized using \eq{Rminus1} for $\Delta=6$ becomes ${\cal R}\times \chi^2_{\rm min}/\nu=1.26$, corresponding to a probability of 0.03, which is acceptable being about a $2\sigma$  effect.  

Again, we have cleaned up the sample. We have rejected 25 datum points which had an average $\Delta\chi^2_i\sim 15$.
We  have demonstrated that:
(1) the model works, and (2) we had large $\delchi$ contributions from the outliers that we were able to Sieve out. 

Inspection of  Table \ref{pipfit} shows that the parameter values effectively do not depend on our choice of cut-off, $\delchimax$, not changing  significantly compared to the predicted parameter errors.   
Another and perhaps better indication of the stability of the Sieve is illustrated in Table \ref{pippredicted}.  Tabulated as a function of $\sqrt s$ are:
\begin{itemize}
\item the predicted total cross sections and $\rho$-values for $\pi^-p$ and $\pi^+p$
\item the errors in their predictions generated by the errors in the fit parameters $c_1,\ c_2$ and $f_+(0)$
\end{itemize}
 for two different values of the cut-off, $\delchimax=4$ and $\delchimax=6$. The predicted cross sections and $\rho$ values for the two values of $\delchimax$  are essentially indistinguishable, again generating  strong confidence in the Sieve technique when used with these four different examples of real-world experimental data.

\begin{figure}[p] 
\begin{center}
\mbox{\hspace{.5in}\epsfig{file=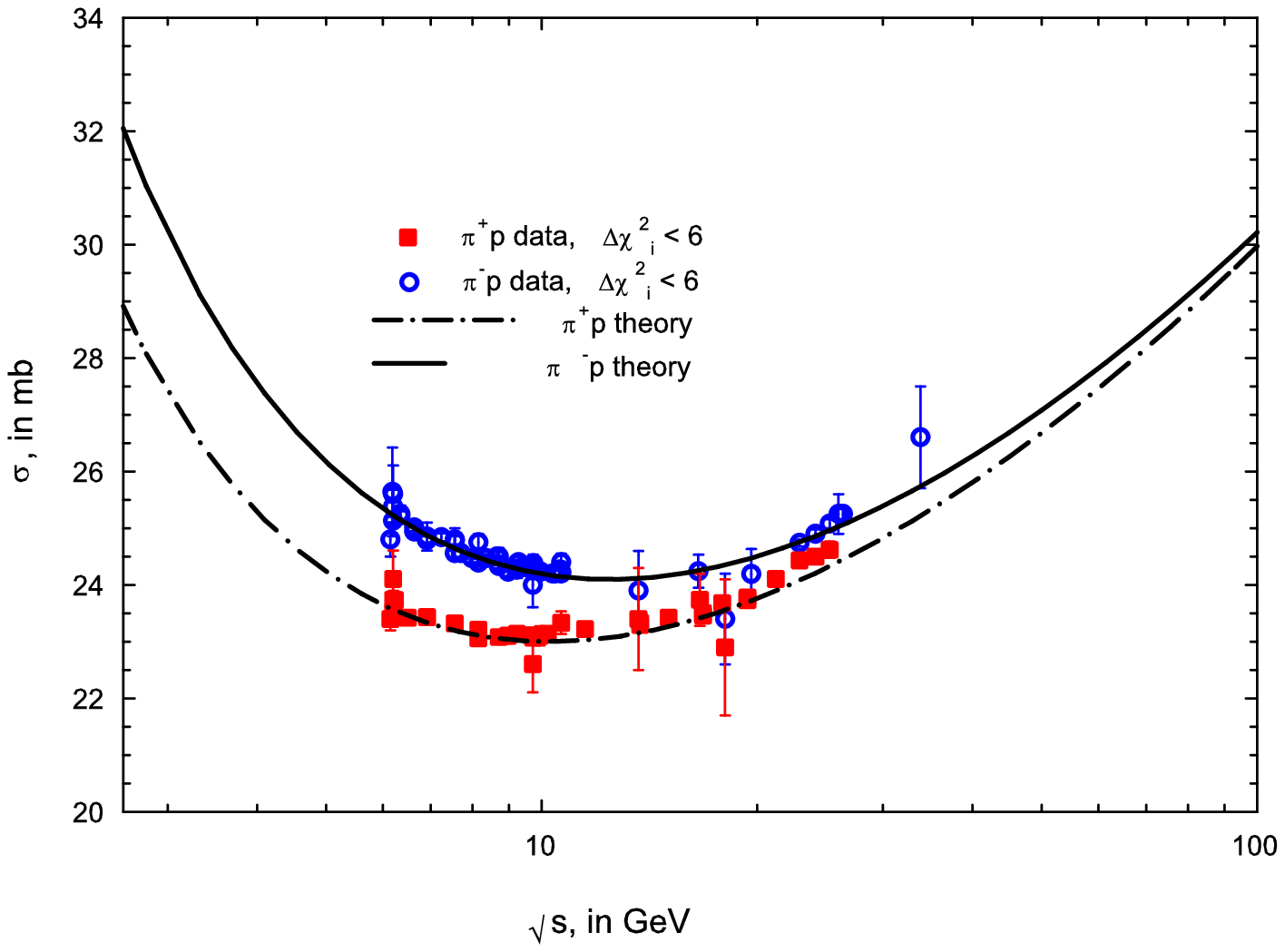,width=4.9in%
,bbllx=0pt,bblly=0pt,bburx=415pt,bbury=300pt,clip=%
}}
\end{center}
\caption[]{ \footnotesize
 The data points shown are the result of screening all of the points of Fig. \ref{pipdg} for those cross section points with $\Delta \chi^2_i<6.$  The open circles are $\sigma_{\pi^-p}$ and the squares are $\sigma_{ \pi^+p}$. The solid line is the theoretical fit to $\sigma_{\pi^-p}$ and the dashed line is the theoretical fit to $\sigma_{\pi^+p}$. 
  }
\label{pisig}
\end{figure}

\begin{figure}[h,t,b] 
\begin{center}
\mbox{\epsfig{file=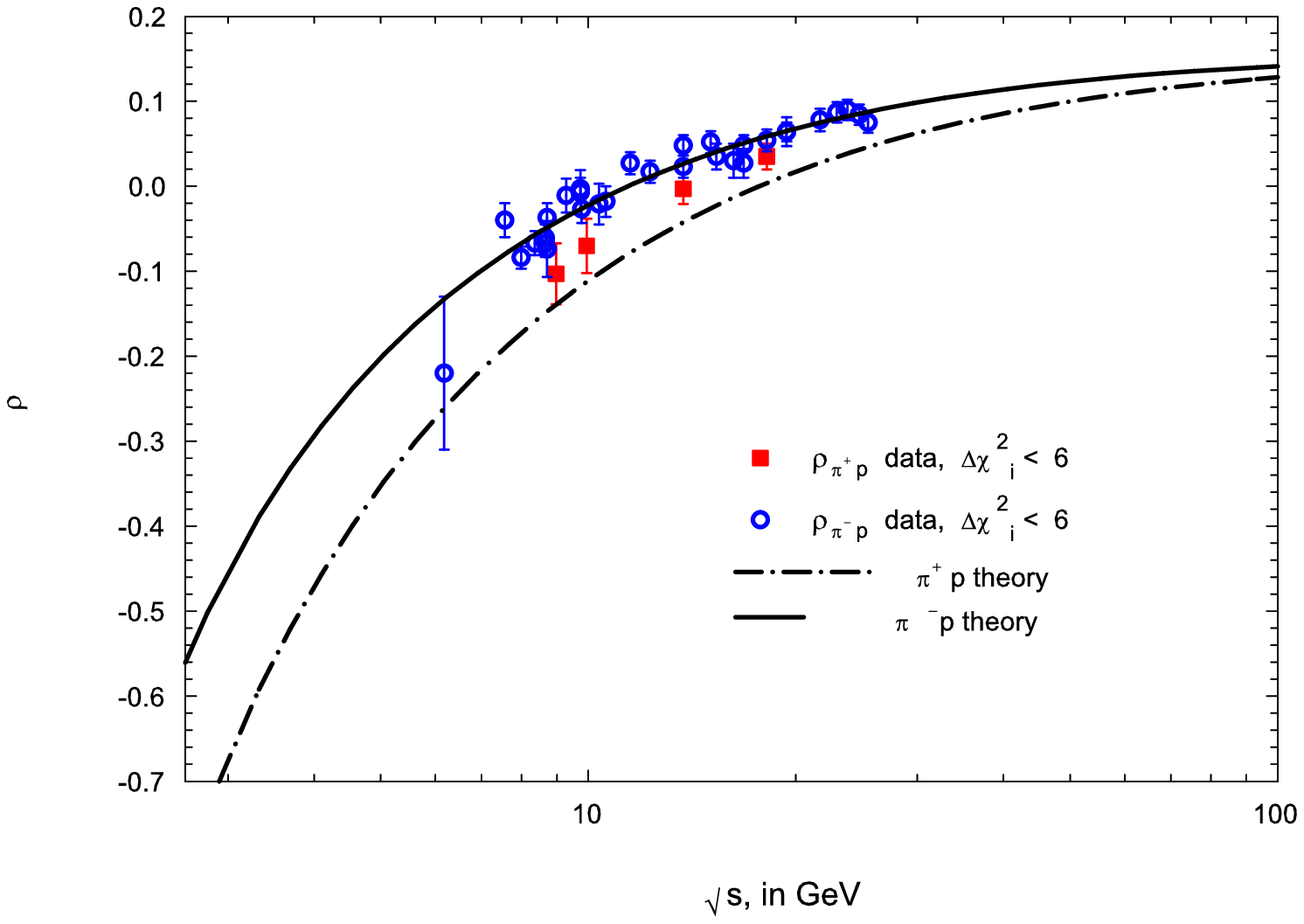,width=5.3in%
,bbllx=90pt,bblly=330pt,bburx=545pt,bbury=641pt,clip=%
}}
\end{center}
\caption[]{ \footnotesize
The data points shown are the result of screening all of the points in Fig. \ref{pirhopdg} for those $\rho$-value points with $\Delta \chi^2_i<6.$  The open circles are $\rho_{\pi^-p}$ and the squares are $\rho_{ \pi^+p}$. The solid line is the theoretical fit to $\rho_{\pi^- p}$ and the dashed line is the theoretical fit to $\rho_{\pi^+ p}$. 
 
  }
\label{pirhosmall}
\end{figure}

The results of applying the Sieve algorithm to the 4 data sets, along with the fitted curves, are graphically shown in Fig. \ref{pisig} for $\sigma_{\pi^-p}$ and $\sigma_{\pi^+p}$ and in Fig. \ref{pirhosmall}  for $\rho_{\pi^-p}$ and $\rho_{\pi^+p}$.  The fits shown are in reasonable agreement with the 155 data points retained by the Sieve.

Again, when we attempted to fit the sifted  data set of 130 points with a $\ln s$ fit, we found $\chi^2_{\rm min}=942.5$, with $\nu=128$, giving $\chi^2/\nu=7.35$, with a probability of $<<10^{-45}$.  Thus, again a $\ln^2 s$ fits well and a $\ln s$ fit is ruled out for the $\pi p$ system.

\section{Comments and conclusions}
We have shown that the Sieve algorithm works well in the case of  backgrounds in the range  of 0  to $\sim 40\%$, {\em i.e.,}  for  extensive computer data that were generated about a straight line, as well as about a constant, and  for a single  event with a 20\% outlier contamination as well as a 13\%``inlier'' contamination,  that was generated about a parabola.  It also works well for the $\sim 13\%$ to 19\% contamination for the eight real-world data sets taken from the Particle Data Group\cite{pdg}. However, the Sieve algorithm is clearly inapplicable in the situation where the outliers (noise) swamps the signal. In that case, nothing can be done.

There are many possible choices for distributions resulting in robust fits. Our particular choice of  minimizing the Lorentzian squared, $\Lambda^2_0(\alphabold;\xbold)\equiv\sum_{i=1}^N\ln\left\{1+0.18\delchisq\right\}$, in order to extract the parameters $\{\alpha_1,\ldots,\alpha_M\}$ needed to apply our Sieve technique seems to be a sensible one for both artificial computer-generated noisy distributions, as well as for real-world experimental data. This statement should not  be interpreted as meaning that real-world data is truly well-approximated  as a Lorentz distribution, but rather,  as demonstrating that   using the Lorentz distribution to get rid of outliers without sensibly affecting the fit parameters works well in the real world.  Next, the choice of filtering out all points with $\delchi>\delchimax$---where $\delchimax$ is as large as possible---is  optimal in both minimizing the loss of good data and maximizing the loss of outliers, resulting in a renormalized ${\cal R}\times\chi^2_{\rm min}/\nu\sim 1$ for both the computer-generated and the real-world sample, as well as  minimizing the distribution widths, and thus, the errors assigned to the parameters.  

In detail, the utilization of the ``Sieved'' sample with  $\delchi<\delchimax$ allows one to 
\begin{itemize}
\item use the {\em unbiased} parameter values found in a $\chi^2$ fit to the truncated sample for the cut $\delchisqmax$, even in the presence of considerable background.
  
\item find the renormalized $\chi^2_{\rm min}/\nu$, {\em i.e.}, ${\cal R}\times\chi^2_{\rm min}/\nu$, where $\cal R$ is the inverse of the factor given in \eq{Rminus1} as a function of $\Delta=\delchisqmax$ and  plotted in Figure \ref{renorm}. 
\item
use the renormalized $\chi^2_{\rm min}/\nu$ to estimate the goodness-of-fit of the model employing  the standard $\chi^2$ probability distribution function.  We thus estimate the probability that  the data set fits the model, allowing one to decide whether to accept or reject the model.  
\item make a robust evaluation of the parameter errors and their correlations, by multiplying the standard covariance matrix $C$ found in the $\chi^2$ fit by the appropriate value of $(r_{\chi^2})^2$ for the cut $\delchimax$.  The value of $r_{\chi^2}$ is given by \eq{rho} and shown in Figure \ref{renorm} as a function of the cut $\delchimax$, where it is called $\Delta$. It ranges from 1 for very large $\Delta$ to $\sim 1.14$ for $\Delta=2$ in \eq{rho}. However, this is not the complete story.    The parameter error is $\sigma=r_{\chi^2}\times \Sigma$ and  we must also take into account the increase in $\Sigma$ due to the cut $\Delta$, which causes the loss of signal points. As shown in Table \ref{table:rchisq} and discussed in detail in Section \ref{section:lessons},  the true loss of accuracy at $\Delta=2$---relative to an unsifted sample of signal data---is the factor $\sim 1.25$. Thus, the algorithm starts failing rapidly for cuts $\Delta$ smaller than 2.
\end{itemize}
In conclusion, the `` Sieve'' algorithm gains its strength  from the combination of making first a $\Lambda^2_0$ fit to get rid of the outliers and then a $\chi^2$ fit to the  sifted data set. By varying the $\delchisqmax$ to suit the data set needs, we easily adapt to the different contaminations of outliers that can be present in real-world experimental data samples. 

Not only do we now have a  robust goodness-of-fit estimate, but we also have also a robust estimate of the parameters and, equally important,  a {\em robust} estimate of their errors and correlations. The phenomenologist can now eliminate the use of possible personal bias and guesswork in ``cleaning up'' a large data set. 
\section{Acknowledgements}
I would like to thank Professor Steven Block of Stanford University for valuable criticism and contributions to this manuscript and  Professor Louis Lyons of Oxford University for many valuable discussions.  Further, I would like to acknowledge the hospitality of the Aspen Center for Physics.
\appendix
\section {Robust Estimation}\label{appendix}
The terminology, ``robust'' statistical estimators\cite{box}, was first introduced to deal with small numbers of data points which have a large departure from the model predictions, {\em i.e.}, outlier points. Later, research on robust estimation\cite{huber,hampel} based on influence functions was carried out. More recently, robust estimations using  regression models\cite{regression} were made---these are inadequate for fitting non-linear models which often are needed in practical  applications.  For example, the fit needed for  \eq{rhopm} is a {\em non-linear} function of the coefficients $c_0,c_1,c_2,\ldots$, since it is the ratio of two linear functions. 
We will discuss one possible technique for handling outlier points in a {\em non-linear} fit when we introduce the Lorentz probability density function in Section \ref{section:lorentz}.
\subsection{Maximum Likelihood Estimates}
Let $P_i$ be the probability density of the $i$th individual measurement, $i=1,\ldots,N$, in the interval $\Delta y$. Then the probability of the total data set is 
\be
{\cal P}=\prod_{i=1}^{N} P_i\Delta y.
\ee 
Let us define the quantity
\be
\Delta \chi^2_i(x_i;\alphabold)\equiv \left(\frac{y_i-y(x_i;\alphabold)}{\sigma_i}\right)^2\label{delchisq},
\ee
where $y_i$ is the measured value at $x_i$, $y(x_i;\alphabold)$ is the expected (theoretical) value from the model under consideration, and $\sigma_i$ is the experimental error of the $i$th measurement. The $M$ model parameters $\alpha_k$ are given by the  $M$-dimensional vector
$
\alphabold =\{\alpha_1,\ldots,\alpha_M\}.
$

$\cal P$ is identified as the likelihood function, which we shall maximize as a function of the parameters $\alphabold=\{\alpha_1,\ldots,\alpha_M\}$.

For the special case where the errors are normally  distributed (Gaussian distribution), we have the likelihood function $\cal P$ given as
\be
{\cal P}=\prod_{i=1}^{N}\left\{\exp\left [-\frac{1}{2}\left(\frac{y_i-y(x_i;\alphabold)}{\sigma_i}\right )^2\right ]\frac{\Delta y}{\sqrt{2\pi}\sigma_i}\right\}=\prod_{i=1}^{N}\left\{\exp\left [-\frac{1}{2}
\Delta \chi^2_i\right ]\frac{\Delta y}{\sqrt{2\pi}\sigma_i}\right\},\label{likelihood}
\ee
Maximizing the likelihood function $\cal P$ in \eq{likelihood} is the same as minimizing the negative logarithm of $\cal P$, namely,
\be
\sum_{i=1}^N  \frac{1}{2}\left(\frac{y_i-y(xi;\alphabold)}{\sigma_i}\right)^2-N\ln\frac{\Delta y}{\sqrt{2\pi}\sigma_i}.
\ee
Since $N$, $\Delta y$ and $\sigma_i$ are constants, after using \eq{delchisq}, this is equivalent to minimizing the quantity
\be
\frac{1}{2}\sum_{i=1}^N\delchisq.\label{2chisq}
\ee

We now define $\chi^2(\alphabold;\xbold)$ as
\be
\chi^2(\alphabold;\xbold)=\sum_{i=1}^N\delchisq,
\ee
where $\xbold\equiv\{x_1,\dots,x_i,\ldots,x_N\}$.

Hence, the $\chi^2$ minimization problem, appropriate to the Gaussian distribution, reduces to
\be
{\rm minimize\ over\ }\alphabold,\quad\quad\chi^2(\alphabold;\xbold)=\sum_{i=1}^N\delchisq
\ee
   for the set of $N$ experimental points at $x_i$ having the value $y_i$ and error $\sigma_i$.
\subsection{Gaussian Distribution}\label{appendix:chisq}
To minimize $\chi^2$, we must solve the (in general, non-linear) set of $M$ equations
\be
\sum_{i=1}^N\frac{1}{\sigma_i}\left(\frac{y_i-y(x_i;\alphabold)}{\sigma_i}\right)\left(\frac{\partial y(x_i;\ldots\alpha_j\ldots)}{\partial\alpha_j}\right)=0,\quad j=1,\ldots,M.\label{chieqns}
\ee
The Gaussian distribution allows a $\chi^2$ minimization routine to return several exceedingly useful statistical quantities.  Firstly, it returns the best-fit parameter space $\alphabold_{\rm min}$. Secondly, the value of $\chi^2_{\rm min}$, when compared to the number of degrees of freedom   ( d.f.$\equiv\nu= N-M$, the number of data points minus the number of fitted parameters) allows one to make  standard estimates of the goodness of the fit of the data set to the model used, using the $\chi^2$ probability distribution function, given in standard texts\cite{nr}, for $\nu$ degrees of freedom.  Further, $C^{-1}$, the $M\times M$ matrix of the partial derivatives at the minimum, given by 
\be
\left[C^{-1}\right]_{jk}=\frac{1}{2}\left(\frac{\partial^2\chi^2}{\partial\alpha_j\partial\alpha_k}\right)_{\alphabold=\alphabold_{\rm min}},
\ee
allows us to compute the standard covariance matrix $C$ for the individual parameters $\alpha_i$, as well as the correlations between $\alpha_j$ and $\alpha_k$\cite{nr}. Thus, when the errors are distributed normally, the $\chi^2$ technique not only gives us the desired parameters $\alphabold_{\rm min}$, but also furnishes  us with  statistically meaningful error estimates of the fitted parameters, along with goodness-of-fit information for the data to the chosen model---very valuable quantities for any model under consideration.
\subsection{Robust Distributions}
We can generalize the maximum likelihood function of \eq{likelihood}, which is  a function of the variable $\frac{y_i-y(x_i; \alphabold)}{\sigma_i}$, as
\be
{\cal P}=\prod_{i=1}^{N}\left\{\exp\left [-\rho\left( \frac{y_i-y(x_i; \alphabold)}{\sigma_i}\right)
\right ]\Delta y\right\},\label{genP}
\ee
where the function $\rho\left( \frac{y_i-y(x_i; \alphabold)}{\sigma_i}\right)$ is the negative logarithm of the probability density. Note that the statistical function $\rho$ used in this Appendix has nothing to do with the $\rho$-value used in \eq{rhopm}. 
Thus, we now have to minimize the generalization of \eq{2chisq}, {\em i.e.,}
\be
{\rm minimize\ over\ }\alphabold,\quad\quad \sum_{i=1}^N\rho\left( \frac{y_i-y(x_i; \alphabold)}{\sigma_i}\right),\label{newmin}
\ee
for the $N$-dimensional vector $\xbold$.

This yields the more general set of $M$ equations
\be
\sum_{i=1}^N\frac{1}{\sigma_i}\psi\left(\frac{y_i-y(x_i;\alphabold)}{\sigma_i}\right)\left(\frac{\partial y(x_i;\ldots\alpha_j\ldots)}{\partial\alpha_j}\right)=0,\quad j=1,\ldots,M,\label{robusteqns}
\ee
where the influence function $\psi(z)$ in \eq{robusteqns} is given by
\be
\psi(z)\equiv\frac{d\beta(z)}{dz},\quad z\equiv\frac{y_i-y(x_i;\alphabold)}{\sigma_i}={\rm sign}(y_i-y(x_i;\alphabold))\times\sqrt{\delchisq}.\label{z}
\ee
Comparison of \eq{robusteqns} with the Gaussian equivalent of \eq{chieqns} shows that
\be
\rho(z)=\frac{1}{2}z^2,\quad \psi(z)=z \quad{\rm (for\ a \ Gaussian \ distribution)}.
\ee
We note that for a Gaussian distribution,
the influence function $w(z)$ for each experimental point $i$ is proportional to $\sqrt{\Delta \chi^2_i}$, the normalized departure of the point from the theoretical value. Thus, the more the departure from the theoretical value, the more ``influence'' the point has in minimizing $\chi^2$. This gives outliers (points with large departures from their theoretical values) unduly large ``influence'' in computing the best vector \mbox{\boldmath $\alpha$}, easily skewing the answer due to the inclusion of these outliers.
\subsection{Lorentz Distribution}\label{section:lorentz}
Consider the normalized Lorentz probability density distribution (also known as the Cauchy distribution or the Breit-Wigner line width distribution), given by 
\begin{eqnarray}
P(z)&=&\frac{\sqrt{\gamma}}{\pi}\frac{1}{1+\gamma z^2},\label{cauchy1}
\end{eqnarray}
where $\gamma$ is a constant whose significance  will be discussed later. 
Using \eq{delchisq} and \eq{z},  we rewrite \eq{cauchy1} in terms of the measurement errors $\sigma_i$ and the  experimental measurements  $y_i$ at $x_i$  as
\begin{eqnarray}
P\left(\frac{y_i-y(x_i;\alphabold)}{\sigma_i}\right)&=&\frac{\sqrt{\gamma}}{\pi}\frac{1}{1+\gamma\left(\frac{y_i-y(x_i;\alphabold)}{\sigma_i}\right)^2}\nonumber\\
&=&\frac{\sqrt{\gamma}}{\pi}\frac{1}{1+\gamma\delchisq}.\label{cauchy}
\end{eqnarray}
It has long tails and therefore is more suitable for robust fits than is the Gaussian distribution.
Taking the negative logarithm of \eq{cauchy} and using it in \eq{newmin}, we see that 
\begin{eqnarray}
\rho(z)&=&\ln\left(1+\gamma z^2\right)=\ln\left\{1+\gamma\delchisq\right\}\quad{\rm and}\nonumber\\
\psi(z)&=&\frac{z}{1+\gamma z^2}
=\frac{{\rm sign}(y_i-y(x_i;\alphabold))\times\sqrt{\delchisq}}{1+\gamma\delchisq}.\label{lorentz}
\end{eqnarray}

In analogy to  $\chi^2$ minimization, we must now minimize $\Lambda^2(\alphabold;\xbold)$, the Lorentzian squared, 
with respect to the parameters $\alphabold$, for a given set of experimental points $\xbold$, {\em i.e.,}
\be
{\rm minimize\ over\ }\alphabold,\quad\quad\Lambda^2(\alphabold;\xbold)\equiv\sum_{i=1}^N\ln\left\{1+\gamma\delchisq\right\}\label{lormin},
\ee
 for the set of $N$ experimental points at $x_i$ having the value $y_i$ and error $\sigma_i$.

We have made extensive computer simulations using Gaussianly generated data (constant and straight line models) which  showed empirically that the choice $\gamma=0.18$ minimized the rms (root mean square) parameter widths found in $\Lambda^2$ minimization.  Further, it gave rms widths that were almost as narrow as those found in $\chi^2$ minimization on the same data.   We will adopt this value of $\gamma$, since it effectively minimizes the width for the  $\Lambda^2$ routine, which we now call $\Lambda^2_0(\alphabold;\xbold)$. 
Thus we select for our robust algorithm, 
\be
{\rm minimize\ over\ }\alphabold,\quad\quad\Lambda^2_0(\alphabold;\xbold)\equiv\sum_{i=1}^N\ln\left\{1+0.18\delchisq\right\}\label{lormin0}.
\ee
An important property of $\Lambda^2_0(\alphabold;\xbold)$ is that it numerically gives the same total $\chi^2_{0_{\rm min}}$ as that found in a $\chi^2$ fit, {\rm i.e.} $\chi^2_0=\sum _{i=1}^N\delchisq$,  where the $\delchisq$ come from the minimization of $\Lambda_0^2$ in \eq{lormin0}, is the same as the $\chi^2_{\rm min}$ found   using a standard $\chi^2$ minimization on the same data.

We note from \eq{lorentz} that the influence function for a point $i$ for small $\sqrt{\Delta \chi^2_i}$ increases proportional to $\sqrt{\Delta \chi^2_i}$ (just like the Gaussian distribution does), whereas for large $\sqrt{\Delta \chi^2_i}$, it {\em decreases} as $1/\sqrt{\Delta \chi^2_i}$. Thus, large outliers have  {\em much less} ``influence'' on the fit than do points close to the model curve---this feature makes $\Lambda^2$ minimization robust. Thus, outliers have  little influence on the choice of the parameters $\alphabold_{\rm min}$ resulting from the minimization of $\Lambda^2_0$, a major consideration for a robust minimization method.

Unlike the minimization of $\chi^2$, the minimization of $\Lambda^2_0$, while yielding the desired  robust estimate of $\alphabold_{\rm min}$,  gives neither parameter error information on $\alphabold_{\rm min}$ nor   a conventional goodness-of-fit. These are major failings, since one has no objective grounds for accepting or rejecting the model.   We will rectify  these shortcomings in the main section of the text, Section \ref{section:sieve}, where we describe the adaptive ``Sieve'' algorithm.  Extensive computer studies, summarized in Section \ref{section:lessons}, demonstrate that use of  this algorithm enables one to make a {\em robust} error estimate of $\alphabold_{\rm min}$, as well as a {\em robust} estimate of the goodness-of-fit of the data to the model.

\newpage
\begin{table}[h,t]                   
%
\def\arraystretch{1.5}            
\begin{tabular}[b]{|c||c|c|c|c||}
\cline{2-5}
	\multicolumn{1}{c|}{}
	&\multicolumn{1}{c|}{$\delchimax=9$ }
	&\multicolumn{1}{c|}{$\delchimax=6$ }
	&\multicolumn{1}{c|}{$\delchimax=4$ }
	&\multicolumn{1}{c||}{$\delchimax=2$ }\\

\hline
$r_{\chi^2, {\rm str.\ line}}$&1.034&1.054&1.098&1.162\\
$r_{\chi^2,{\rm  constant}}$&1.00&1.05&1.088&1.108\\
\hline
average&1.018&1.052&1.093&1.148\\
\hline\hline
$<\chi^2_{\rm min}>/\nu$&&&&\\
str. line&0.974&0.901&0.774&0.508\\
constant&0.973&0.902&0.774&0.507\\
\hline
average&0.973&0.901&0.774&0.507\\
${\cal R}^{-1}$&0.9733&0.9013&0.7737&0.5074\\
\hline\hline
$S.F.$&0.9973&0.9857&0.9545&0.8427\\
\hline\hline
$\sigma/\sigma_0$&1.02&1.06&1.19&1.25\\
\hline\hline
\end{tabular}
     \caption{\protect\small Results for $r_{\chi^2}=\sigma/\Sigma$, the ratio of the rms width to $\Sigma$, the error for the $\chi^2$ fit;  $<\chi^2_{\rm min}>/\nu$, for both the straight line case  and the constant  case;  $\sigma/\sigma_0$, the ratio of the rms width (error) of the parameter relative to what the error would be if the sample were not truncated, {\em i.e.}, the total loss of accuracy due to truncation, as  functions of the cut $\delchimax$.   The average results for $r_{\chi^2}$ and $<\chi^2_{\rm min}>/\nu$ are graphically shown in Fig. \ref{renorm}.
See Sections \ref{section:width}, \ref{section:constant} and \ref{section:lessons} for details. The theoretical values for the renormalization factor ${\cal R}^{-1}$ are from \eq{Rminus1} and the survival fractions $S.F.$ are from \eq{SF}.  See Section \ref{section:lessons} for a discussion of the error-broadening factor $\sigma/\sigma_0$.
\label{table:rchisq}
}
\end{table}
\def\arraystretch{1}  

\begin{table}[h,t]                   
%
\def\arraystretch{1.5}            
\begin{tabular}[b]{|l||c||c||c||}
\hline
\multicolumn{1}{|c||}{Fitted}
      &\multicolumn{3}{|c|}{$\delchimax$}\\ 
\cline{2-4}
	\multicolumn{1}{|c||}{Parameters}
      &\multicolumn{1}{c||}{4}
      &\multicolumn{1}{c||}{6}&\multicolumn{1}{c||}{9}\\
      \hline\hline
      $c_1$\ \ \   (mb)&$-1.452\pm0.066$&$-1.448\pm0.066$ &$-1.423\pm0.065$\\ 
	$c_2$\ \ \ \   (mb)&$0.2828\pm0.0061$&$0.2825\pm0.0060$&$0.2801\pm0.0059$\\
	$f(0)$ (mb GeV)&$-0.065\pm0.56$&$-0.020\pm0.56$&$-0.065\pm0.56$\\
     	\hline
	\hline
	$\chi^2_{\rm min}$&142.8&182.8&217.9\\
	$\nu$ (d.f).&182&190&195\\
\hline
	${\cal R}\times\chi^2_{\rm min}/\nu$&1.014&1.067&1.143\\
\hline
\end{tabular}
     \caption{\protect\small The fitted results for a 3-parameter fit to the total cross sections and $\rho$-values for $pp$ and $\bar pp$ scattering. The renormalized $\chi^2/\nu_{\rm min}$,  taking into account the effects of the $\delchimax$ cut, is given in the row  labeled ${\cal R}\times\chi^2_{\rm min}/\nu$. \label{ppfit}}
\end{table}
\def\arraystretch{1}  
\begin{table}[h,t]                   
%
\def\arraystretch{1.5}            
\begin{tabular}[b]{|l||c|c|c|c||c|c|c|c||c|c|c|c||}
\hline
\multicolumn{1}{|c||}{$\sqrt s$}
      &\multicolumn{4}{|c||}{$\delchimax=4$} &\multicolumn{4}{|c||}{$\delchimax=6$}&\multicolumn{4}{|c||}{Predicted Error}\\ 
\cline{2-13}
	\multicolumn{1}{|c||}{(GeV)}
      &\multicolumn{1}{c|}{$\sigma_{\bar pp}$}
      &\multicolumn{1}{c|}{$\sigma_{ pp}$}&\multicolumn{1}{c|}{$\rho_{\bar pp}$}&\multicolumn{1}{c||}{$\rho_{pp}$}&\multicolumn{1}{c|}{$\sigma_{\bar pp}$}&\multicolumn{1}{c|}{$\sigma_{ pp}$}&\multicolumn{1}{c|}{$\rho_{\bar pp}$}&\multicolumn{1}{c||}{$\rho_{pp}$}&\multicolumn{1}{c|}{$\sigma_{\bar pp}$}&\multicolumn{1}{c|}{$\sigma_{ pp}$}&\multicolumn{1}{c|}{$\rho_{\bar pp}$}&\multicolumn{1}{c||}{$\rho_{pp}$}\\
      \hline\hline
      10&43.77&38.34 &-0.0368&-0.1501&43.77&38.33 &-0.0365&-0.1498&.01&.01&.003&.004\\    
	100&46.61&46.25 &0.1083&0.1031&46.61&46.25 &0.1082&0.1031&.08&.08&.001&.001\\    
	540&60.87&60.82 &0.1368&0.1363&60.86&60.81 &0.1367&0.1362&.28&.28&.001&.001\\    
	1800&75.30&75.29 &0.1396&0.1395&75.28&75.27 &0.1396&0.1395&.50&.50&.001&.001\\    
	14000&107.6&107.6 &0.1318&0.1318&107.5&107.5 &0.1318&0.1318&1.0&1.0&.001&.001\\    
\hline
   
\end{tabular}
     \caption{\protect\small The predicted results for $\sigma_{\bar pp}$, $\sigma_{pp}$, $\rho_{\bar pp}$ and $\rho_{pp}$, together with their errors, as a function of $\sqrt s$, the cms energy in GeV, for  $\delchimax=4$ and $\delchimax=6$ . The cross sections and their errors are in mb. The predicted errors are those found from a standard $\chi^2$ analysis.
\label{predicted}}
\def\arraystretch{1}  
\end{table}
\begin{table}[h,t]                   
%
\def\arraystretch{1.5}            
\begin{tabular}[b]{|l||c||c||c||}
\hline
\multicolumn{1}{|c||}{Fitted}
      &\multicolumn{3}{|c|}{$\delchimax$}\\ 
\cline{2-4}
	\multicolumn{1}{|c||}{Parameters}
      &\multicolumn{1}{c||}{4}
      &\multicolumn{1}{c||}{6}&\multicolumn{1}{c||}{9}\\
      \hline\hline
      $c_1$\ \ \   (mb)&$-0.895\pm0.11$&$-0.921\pm0.11$ &$-0.982\pm0.10$\\ 
	$c_2$\ \ \ \   (mb)&$0.174\pm0.0083$&$0.177\pm0.0081$&$0.182\pm0.0075$\\
	$f(0)$ (mb GeV)&$-2.281\pm0.34$&$-2.307\pm0.34$&$-2.327\pm0.34$\\
     	\hline
	\hline
	$\chi^2_{\rm min}$&128.7&148.1&204.4\\
	$\nu$ (d.f).&122&127&135\\
\hline
	${\cal R}\times\chi^2_{\rm min}/\nu$&1.364&1.293&1.556\\
\hline
\end{tabular}
     \caption{\protect\small The fitted results for a 3-parameter fit to the total cross sections and $\rho$-values for $\pi^+p$ and $\pi^-p$ scattering. The renormalized $\chi^2/\nu_{\rm min}$,  taking into account the effects of the $\delchimax$ cut, is given in the row  labeled ${\cal R}\times\chi^2_{\rm min}/\nu$. \label{pipfit}}
\end{table}
\def\arraystretch{1}  

\begin{table}[h,t]                   
%
\def\arraystretch{1.5}            
\begin{tabular}[b]{|l||c|c|c|c||c|c|c|c||c|c|c|c||}
\hline
\multicolumn{1}{|c||}{$\sqrt s$}
      &\multicolumn{4}{|c||}{$\delchimax=4$} &\multicolumn{4}{|c||}{$\delchimax=6$}&\multicolumn{4}{|c||}{Predicted Error}\\ 
\cline{2-13}
	\multicolumn{1}{|c||}{(GeV)}
      &\multicolumn{1}{c|}{$\sigma_{\pi^-p}$ }
      &\multicolumn{1}{c|}{$\sigma_{\pi^+p}$}&\multicolumn{1}{c|}{$\rho_{\pi^-p}$}
&\multicolumn{1}{c||}{$\rho_{\pi^+p}$ }
      &\multicolumn{1}{c|}{$\sigma_{\pi^-p}$}&\multicolumn{1}{c|}{$\sigma_{\pi^+p}$}&\multicolumn{1}{c|}{$\rho_{\pi^-p}$}
&\multicolumn{1}{c||}{$\rho_{\pi^+p}$}

&\multicolumn{1}{c|}{$\sigma_{\pi^-p}$}&\multicolumn{1}{c|}{$\sigma_{ \pi^+p}$}&\multicolumn{1}{c|}{$\rho_{\pi^-p}$}&\multicolumn{1}{c||}{$\rho_{\pi^+p}$}\\
      \hline\hline
      6&25.40&23.70 &-0.1391&-0.2704&25.40&23.70 &-0.1396&-0.2708&.01&.01&.009&.010\\    
	15&24.26&23.35 &0.0392&-0.0248&24.27&23.36 &0.0396&-0.0243&.01&.01&.002&.002\\    
	23.5&24.92&24.25 &0.0827&0.0385&24.94&24.27 &0.0833&0.0393&.02&.02&.002&.002\\    
	62.5&28.15&27.81 &0.1309&0.1117&28.20&27.86 &0.1318&0.1127&.09&.09&.003&.003\\    	   
\hline   
\end{tabular}
     \caption{\protect\small The predicted results for $\sigma_{\pi^-p}$, $\sigma_{\pi^+p}$,   $\rho_{\pi^-p}$ and $\rho_{\pi^+p}$, together with their errors, as a function of $\sqrt s$, the cms energy in GeV, for  $\delchimax=4$ and $\delchimax=6$ . The cross sections and their errors are in mb. The predicted errors are those found from a standard $\chi^2$ analysis.
\label{pippredicted}}
\def\arraystretch{1}  
\end{table}

\newpage

\begin{figure}[h,t,b] 
\begin{center}
\mbox{\epsfig{file=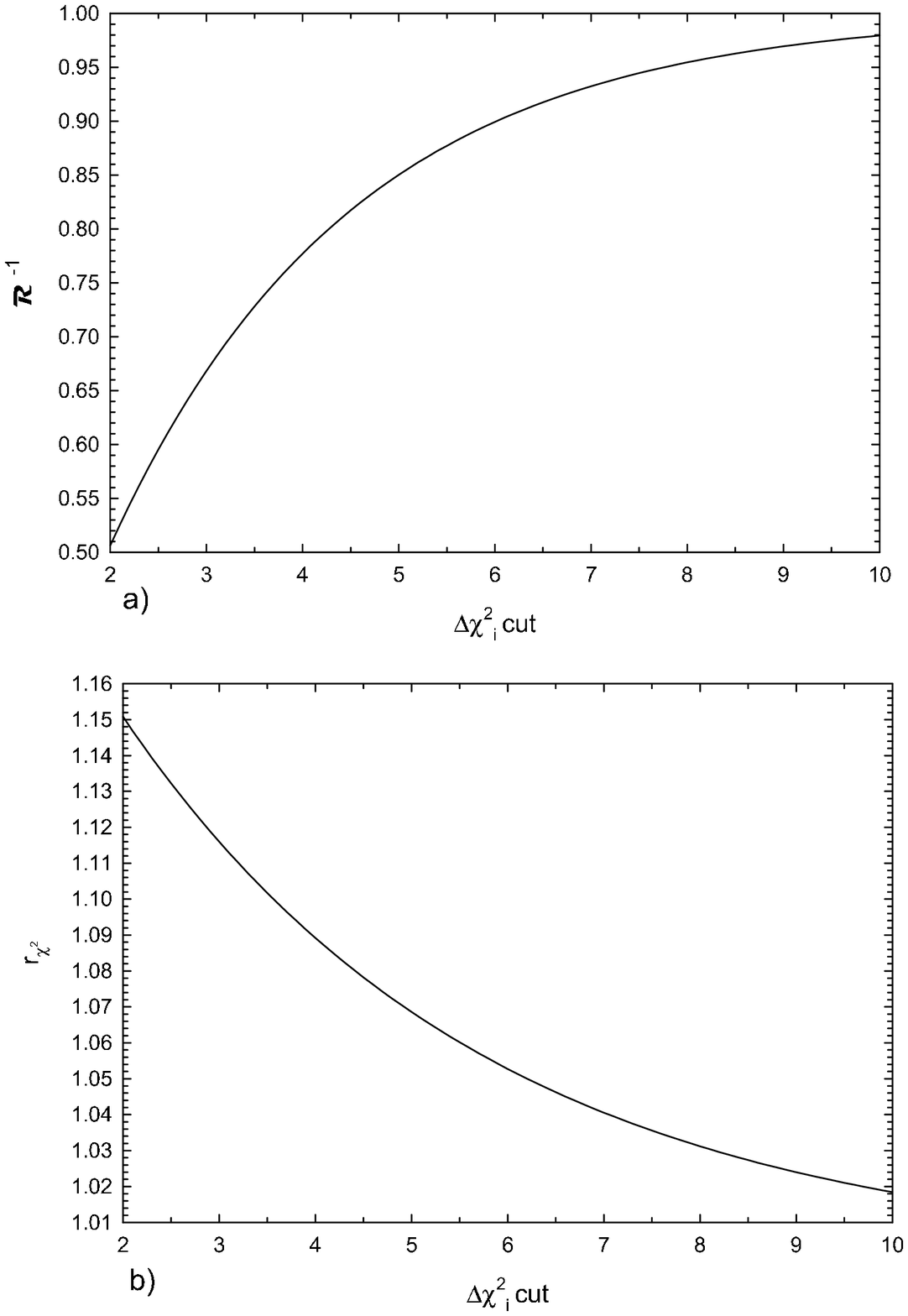,width=5.5in%
,bbllx=65pt,bblly=135pt,bburx=510pt,bbury=775pt,clip=%
}}
\end{center}
\caption[]{ \footnotesize 
a) A plot of \eq{Rminus1}:  ${\cal R}^{-1}$, the reciprocal of the factor  that multiplies $\chi^2_{\rm min}/\nu$ found in the $\chi^2$ fit to the sifted data set  {\em vs.} $\delchi$ cut, the $\delchimax$ cut.
\mbox{\ \ } b) A plot of \eq{rho}: $r_{\chi^2}$, the  factor whose square multiplies the covariant matrix found in the $\chi^2$ fit to the sifted data set  {\em vs.} $\delchi$, the $\chi^2$ cut.  
See Sections \ref{section:width}, \ref{section:constant} and \ref{section:lessons} for details. In  \eq{rho} and \eq{Rminus1}, the $\delchi$ cut is called $\Delta$.
  }
\label{renorm}
\end{figure}


\begin{thebibliography}{99}
\bibitem{random} 
In this context, a random distribution means a uniform distribution between $a$ and $b$, generated by a random number generator that has a flat output between 0 and 1. A normally distributed (Gaussian) distribution means using a Gaussian random number generator that has as its output random numbers $y_i$ distributed normally about $\bar y$, with a probability density $\frac{1}{\sqrt 2 \pi}\exp-\frac{1}{2}\left(\frac{\bar y-y_i}{\sigma_i}\right)^2$, where $\sigma_i$ represents the error (standard deviation) of the point $y_i$.
\bibitem{subtle} The fact that $r_{\chi^2}$ is greater than 1 is counter-intuitive.  Consider the case of generating a Gaussian distribution with unit variance about the value $y=0$. If we were to define $\delchi\equiv  (y_i-0)^2=y_i^2$, with $\Delta$ being the cut $\delchimax$,  then the truncated differential probability distribution would be $P(x)=\frac{1}{\sqrt{2\pi}}\exp({-x^2/2})$ for $-\sqrt{\Delta}\le x \le +\sqrt{\Delta}$, whose rms value clearly is {\em less than} 1---after all, this distribution is truncated compared to its parent Gaussian distribution.  However, that is not what we are doing.  What we do is to first make a robust fit  to each  untruncated event that was Gaussianly generated  with unit variance about  the mean value zero.  For every  event we then find the value $y_0$, its best fit parameter, which, although close to zero with a mean of zero, is {\em non-zero}.  In order to obtain the truncated event whose width we sample with the {\em next} $\chi^2$ fit, we use $\delchi\equiv(y_i-y_0)^2$.  It is the jitter in $y_0$ about zero that is responsible for the rms width becoming greater than 1. This result is true even if  the first fit to the untruncated data were a $\chi^2$ fit.
\bibitem{physics}
In deriving these equations, we have employed real analytic amplitudes derived using unitarity, analyticity, crossing symmetry, Regge theory and the Froissart bound.
\end{thebibliography}
\end{document}